\documentclass[showpacs,amsmath,amssymb, nobibnotes, aps, prl,showkeys, floatfix]{revtex4}
\usepackage{graphicx}
\usepackage{dcolumn}
\usepackage{bm}
\usepackage{docs}
\usepackage{amssymb}
\usepackage{amsmath}
\usepackage{epsfig}
\expandafter\ifx\csname package@font\endcsname\relax\else
 \expandafter\expandafter
 \expandafter\usepackage
 \expandafter\expandafter
 \expandafter{\csname package@font\endcsname}%
\fi

\newcommand{\ltsima} {$\; \buildrel < \over \sim \;$}
\newcommand{\gtsima} {$\; \buildrel > \over \sim \;$}
\newcommand{\lta} {\lower.5ex\hbox{\ltsima}}
\newcommand{\gta} {\lower.5ex\hbox{\gtsima}}

\newcommand{\lsim}{\raisebox{-.4ex}{$\stackrel{<}{\scriptstyle \sim}$}}
\newcommand{\gsim}{\raisebox{-.4ex}{$\stackrel{>}{\scriptstyle \sim}$}}

\newcommand{\RNum}[1]{\uppercase\expandafter{\romannumeral #1\relax}}

\begin{document}

\title[Accretion-outflow coupling in black hole systems ]
{A 2.5-dimensional viscous, resistive, advective magnetized accretion-outflow coupling in black hole systems: A higher order polynomial approximation. \RNum{1}}

\author{Shubhrangshu Ghosh  \thanks{E-mail: sghosh@jcbose.ac.in} }

\affiliation{ Center for Astroparticle Physics and Space Science, Department of Physics, Bose Institute, Block EN, Sector V, Salt Lake, 
Kolkata, India 700091 }

\begin{abstract}

The correlated and coupled dynamics of accretion and outflow around 
black holes (BHs)
are essentially governed by the fundamental laws of conservation as 
outflow extracts matter, momentum and energy from the accretion region. Here we analyzed a robust form of 2.5-dimensional 
viscous, resistive, advective magnetized accretion-outflow coupling in BH systems,  
in the mean field magnetohydrodynamical (MHD) regime. We solve the complete set of coupled  
MHD conservation equations self-consistently, 
through invoking a generalized 
polynomial expansion in two dimensions. We perform a critical analysis of accretion-outflow 
region and provide a complete quasi-analytical family of solutions for advective flows. We obtain the physical plausible outflow solutions at high 
turbulent viscosity parameter $\alpha \, (\gsim \, 0.3)$, and at a reduced scale-height, 
as magnetic stresses compress or squeeze the flow region. We found that the value of the 
large-scale poloidal magnetic field $\bar B_P$ is enhanced with increasing 
geometrical thickness of the accretion flow. On the other hand differential magnetic torque ($-r^2 \bar B_{\varphi} \bar B_z$) 
increases with the increase in $\dot M$. $\bar B_P$, $-r^2 \bar B_{\varphi} \bar B_z$
as well as the plasma beta $\beta_P$ get strongly augmented with the increase in the value of $\alpha$, 
enhancing the transport of vertical flux outwards. 
Our solutions indicate that magnetocentrifugal 
acceleration plausibly plays a dominant role in effusing out plasma from the 
radial accretion flow in moderately advective paradigm which are more centrifugally dominated, however in strongly advective 
paradigm it is likely that the thermal pressure gradient would play a more contributory role in the vertical transport of the 
plasma.




\end{abstract}

\maketitle	

\section{Introduction}

Outflows and jets are ubiquitous in nature. They are observed both in 
local universe, mostly in black hole (BH) X-ray binaries (BHXRBs) which 
are believed to harbour stellar mass BHs called microquasars [1], as well 
as in powerful extragalactic radio sources [2] where well-collimated 
outflows or jets emerge continuously from the nuclear region of the host 
active galaxies (AGNs) or quasars harbouring supermassive BHs. 
The accreting hot plasma around BHs powered by extreme gravity of 
the central object results in the formation of outflow/jet which 
extracts mass, angular momentum and energy from the inner regions of 
the accretion flow. The outflows in microquasars are observed only in low-hard 
state of BHXRBs [3,4] which are radiatively inefficient.  
Radiatively inefficient accretion flows (RIAFs) are hot gas pressure dominated 
systems which are geometrically thick $(h(r)/r \, \gsim \, 0.1)$ and optically thin, where 
$h(r)$ is the 
scale-height of the accretion region. The dynamics of the 
flow is thus strongly sub-Keplerian and advection dominated [5,6]. RIAFs 
occur when the mass accretion rate $\dot M$ 
is very low (presumably with $\dot M \, \lsim \, 10^{-3} \, \dot M_{\rm Edd} $), where 
$\dot M_{\rm Edd}$ is the Eddington accretion rate or the accretion rate 
corresponding to the Eddington luminosity. Outflows/jets are not likely observed  
in high-soft state of BHXRBs ([4] and references therein), which are believed to be powered by 
geometrically thin and optically thick radiation pressure dominated Keplerian accretion disk [7]. 

Theoretically speaking, it has been argued [5,8,9,10]   
that a geometrically thick advective accretion flow has a strong 
tendency to drive bipolar outflows due to high thermal energy 
content of the hot gas. They may be additionally propitious to 
propel outflows/jets because its vertical thick 
structure enhances the large-scale poloidal component 
of the magnetic field, which plays a critical role in 
launching strong and collimated outflows [11]. 

Apart from low-hard 
state of BHXRBs which power jets, at the other end of the spectrum, strong outflows and jets are observed in low 
excitation radio galaxies (LERGs) harbouring supermassive BHs. LERGs, a more generalization of low luminous AGNs (LLAGNs) seem to be 
accreting gaseous plasma directly from the hot X-ray emitting phase of interstellar medium (ISM) or from the 
hot X-ray halos surrounding the galaxy 
or from the hot phase of the intergalactic medium (IGM) quasi-spherically in a radiatively 
inefficient mode with near Bondi rate [12]. LERGs thus resemble low-hard state of BRXRBs having 
geometrically thick and optically thin gas pressure dominated strongly advective 
quasi-spherical accretion flow, accreting hot gas at a high sub-Eddington accretion 
rate (presumably with $\dot M \, \lsim \, 10^{-3} \, \dot M_{\rm Edd} $).  
This strongly advective radiatively inefficient accretion paradigm (RIAF) or hot mode accretion having considerable 
geometrical thickness, is more prone to emanate outflows/jets and is very conducive to propel matter 
vertically outwards out of the accreting region. 

However, with the increase in $\dot M$ as 
$10^{-3} \dot M_{\rm Edd} << \dot M \, \lsim \, 10^{-2} \, \dot M_{\rm Edd}$, the flow tends to 
be more centrifugally dominated and become moderately advective, with the central BH accreting relatively cold gas as compared to 
the hot mode accretion. Incidentally the moderately advective accretion 
flow does not occur in a radiatively inefficient 
mode, but where considerable amount of both gas and 
radiation pressure seem to be present in the system, rendering the flow to 
have a moderate optical depth. Geometrically, the inner advective region would then be 
relatively thinner than that correspond to RIAFs. The moderately advective 
accretion paradigm may also be susceptible to eject outflows. 
The difference between this paradigm and with the RIAF, however, may lie in the acceleration mechanisms to eject 
bipolar outflows and jets, which we would eventually investigate in this study. Nonetheless, with the increase in 
$\dot M$ as the flow tends to become more rotationally/centrifugally dominated with lesser geometrical 
thickness, the efficacy of the disk to eject outflows 
diminishes. Beyond $\dot M > 10^{-2} \, \dot M_{\rm Edd}$, the flow would eventually tend towards 
Keplerian nature ([13]; paper \RNum{2} (in preparation)). Geometrically thin Keplerian accretion disk 
plausibly fails to account for the launching and acceleration of outflows 
and jets ([8,9,10]; also see [14] for further discussion). 

Extensive work has been pursued on the origin of outflow/jet, since the 
seminal work of Blandford \& Payne [15] in studying accretion powered hydromagnetic outflows, which we 
like to focus upon in this study. Physical 
understanding of the accretion powered hydromagnetic outflows have either 
been performed in stationary 
self-­similar approximation [15,16,17] in 
quasi-analytical regime to demonstrate the importance of poloidal component of the magnetic field to 
launch outflowing matter from the Keplerian accretion disk, or through magnetohydrodynamic (MHD) simulations 
in both nonrelativistic as well as in relativistic 
regimes (For details, see the introduction 
in [8,10] and references therein, also see [18] and references therein). 
In most of these studies the authors remain focused mainly on the 
launching of outflows/jets from the geometrically thin Keplerian disk. The formation of the 
accretion powered outflow and 
jet is directly related to the efficacy of extraction of angular 
momentum and energy from the magnetized accretion flow. However, the 
exact mechanism by which the radial accretion is diverted into strong 
outflows and plausible jets still remains theoretically 
elusive. Notwithstanding, jet launching is completely a 
magnetohydrodynamic process. Accreting material diffuses across 
magnetic field lines threading the accretion region, is then lifted 
upwards by MHD forces which then 
couples to the field and becomes accelerated magnetocentrifugally. However, if the accreting 
system is strongly gas pressure dominated, it may happen that the gas pressure 
gradient would play a more contributory role to lift the plasma vertically outwards 
along with the help of magnetic 
forces. In addition, turbulent reconnection of magnetic field lines may lead to 
flux annihilation [19]. Magnetic energy 
dissipates through turbulent magnetic reconnection, may also power 
the outflow/jet emission ([20] and references therein). 

Most of the studies of accretion disk and outflow/jet have evolved 
separately, assuming these two to be apparently dissimilar objects. In 
the light of both deeper theoretical understanding and observational 
inferences [21,22,3] it is evident that the dynamics of 
outflow and the underlying 
accretion are strongly correlated (for details see [8,9,10] and references 
therein). Outflows and jets observed in AGNs and XRBs
can only originate in an accretion powered system, where the
accreting plasma around gravitating objects like BHs acts as a source, whereas outflow and then jet acts as 
one of the possible sinks [10]. The implicit coupling between accretion and 
outflow is then essentially governed by conservation laws; conservation 
of matter, momentum and energy. The outflowing matter carries away mass, angular momentum 
and energy extracted from the accreting plasma [16]. 
We do not intend to investigate the 
physics of jet formation and its launching mechanism which is altogether 
a different field of research, but would like to focus entirely on the 
inter-correlating dynamics of the accretion and outflow within the coupled 
accretion-outflow region, and the conditions/criteria for jet launching. Any proper 
understanding of the dynamics and the conditions of jet launching should necessarily 
require the robust understanding of the dynamics of the magnetized advective accretion region coupled to 
outflow, governed explicitly by the conservation laws. The relevant dynamical 
solutions at the accretion-outflow coupled surface (the surface from where outflow decouples from radially inward accretion flow) 
would then act as boundary conditions at the base of the jet. 

Accretion disk-outflow/jet coupling has been studied on number of occasions from purely observational angle, 
in accretion powered systems [22,23]. From theoretical perspective, few self-similar studies have been attempted 
in the context to accretion-outflow coupling, both in non magnetized and in magnetized regime (see [18] and 
references therein; [8,9,10,24,25,26]). Notwithstanding, theoretically, it is still difficult to construct reasonably satisfactory and 
definitive model of magnetized accretion-outflow/jet coupled region, owing to the complicated geometry and 
inconclusive understanding of the inflow-outflow coupled region. 
On the other hand, few simulations on disk-outflow/jet coupling have also been cultivated [27,28,29]. In these
simulations how the matter gets deflected from
the equatorial plane has been studied largely in the Keplerian regime. 
Casse and Keppens [30] performed an advective, resistive MHD simulation of 
accretion-ejection structure with the inclusion of the energy 
equation, however neglecting the viscosity and radiative loss from the system. Nonetheless, it is still difficult 
to simultaneously simulate the accretion and the outflow regions because 
the time scales of the accretion and outflow are in general very different. 

In the present work, we endeavour to develop a robust viscous, resistive and advective 
MHD accretion-induced outflow model in the 2.5 dimensional regime, in the mean field MHD 
approximation, in the context to accretion powered hydromagnetic outflows/jets, focusing entirely on the 
inter-correlating dynamics of the accretion and outflow within the coupled 
accretion-outflow region, without aspiring to explore the mechanism of launching and ejection of outflows and jets. 
A complete 2.5-dimensional viscous, 
resistive, advective global MHD numerical solution of such a system is left for 
further work, and which is beyond the scope at present; we confine our treatment to quasi-analytical/quasi-numerical 
power law self-similarity (e.g., [5]) in a quasi-stationary configuration,  
by upholding the conservation equations. All physical quantities are scaled as powers in $r$ and $z$ according to their
dimensions, in the limit of higher order polynomial expansion. We will perform a critical analysis of accretion-outflow 
region and provide a complete quasi-analytical family of solutions. Although the quasi-analytical self-similar solutions are 
approximate, they, however, can provide a strong physical intuitive picture of accretion 
dynamics coupled with the outflow, as well as physical criteria/conditions to eject outflows/jets. 
In the next 
section, we present the formulation of our model. \S \RNum{3}
describes the quasi-analytical/quasi-numerical procedure to solve the model equations 
of the accretion-induced outflow. In \S \RNum{4}, we evaluate 
the coefficients of our 
self-similar solutions and analyze them. In \S \RNum{5}, we 
investigate the nature and behaviour of the family of solutions for 
accretion-induced outflow within the bounded accretion coupled outflow 
region. Finally, we 
end up in \S \RNum{6} with a summary and discussion.  

\section{2.5-dimensional advective accretion-outflow coupling in the mean-field MHD regime}

We formulate the accretion-outflow coupled model by considering a 
2.5-dimensional viscous, resistive, advective accretion flow geometry in the mean field 
MHD regime as strong outflows/jets are more likely to eject from a 
geometrically thick, advective region of the accretion 
flow. The vertical flow is explicitly included 
in the system. We adopt the cylindrical coordinate system ($r,\varphi,z$) to 
describe a quasi-stationary, mean axisymmetric accretion flow. As we have incorporated 
outflow in our system, within the accretion-outflow coupled region in advective 
paradigm, all the 
dynamical flow variables; namely, radial velocity ($v_r$), azimuthal 
velocity ($v_{\varphi}$), specific angular momentum ($\lambda$), vertical velocity 
or outflow velocity ($v_z$), isothermal sound speed ($c_s$), 
mass density ($\rho$), thermal pressure ($P$) and magnetic field 
components ($B_r, B_z, B_{\varphi}$) vary in both $r$ and $z$. The dynamical equations are vertically integrated over an 
arbitrary scale height $h(r)$ from $-h$ to $+h$. Here $h(r)$ is not a 
hydrostatic disk-scale height but a photospheric surface within which 
accretion and outflow are coupled. Above $h(r)$, the outflow decouples 
from the accretion flow, gets further accelerated in the hot nonthermal 
magnetized corona and finally forms a relativistic jet. We focus on this 
accretion-outflow coupled region within which the flow is mostly 
bounded. We only include the $r\varphi$ component 
of turbulent stress, which is responsible for radial 
transport of angular momentum outwards (angular momentum get 
transported due to the diffusion of 
turbulent eddies). Vertical transport of angular momentum 
occurs mainly through large-scale magnetic stresses, where the outflowing matter 
magnetically extracts or removes angular momentum. The accreting mass is assumed to be much less 
as compared to that of the central object, and hence the flow is 
not self-gravitating. As the accretion flow is inherently 
turbulent, we express all the dynamical variables in mean and 
fluctuating part. Microscopic viscosity and resistivity  are 
neglected compared to the large-scale diffusion of 
turbulent eddies. Assuming that the fluctuation in the density is 
very less; in the limit of Boussinesq approximation, the generic ansatz follows as  
\begin{eqnarray}
F = \bar F + F^{\prime}, \, \, \, \, \, \, 
\overline{F^{\prime}} = 0, \, \, \, \, \, \,  
\frac{\rho^{\prime}}
{\bar{\rho}} << 1 \, , 
\label{1}
\end{eqnarray}
where $\bar F$ represents the mean flow which is either a 
time average or an ensemble average. $F^{\prime}$ is then the  
fluctuation corresponding to that variable. The turbulence is defined 
in terms of mean Reynolds and Maxwell stress described through correlations given by
\begin{eqnarray}
{\bar t}_{ij} = \overline{t^{\mathcal R}_{ij}} \, + \, \overline{t^{\mathcal M}_{ij}} \, \Rightarrow \, -\left[\overline{\rho v^{\prime}_{i} v^{\prime}_{j}} - 
\Big(\overline{\frac{B^{\prime}_{i} B^{\prime}_{j}}{4 \pi} - \delta_{ij} \frac{{B^{\prime}}^2}{8 \pi}}\Big) \right] \, , 
\label{2}
\end{eqnarray}
where, ${\bar t}_{ij}$ is the net turbulent stress. Statistical averaging of MHD equations generates large number of turbulent correlation terms. In the present study, we restrict to first order turbulent correlation and neglect second order 
and higher order correlation terms. The Reynolds and 
Maxwell stresses are commonly parameterized in terms of kinetic and 
magnetic turbulent viscosities $\nu^{\mathcal R}_{ij}$ and 
$\nu^{\mathcal M}_{ij}$ respectively as 
\begin{eqnarray}
\overline{\rho v^{\prime}_{i} v^{\prime}_{j}} = -\nu^{\mathcal R}_{ij} \, {\bar \rho} \, \overline{ s_{ij}} \, ,
\label{3}
\end{eqnarray}
\begin{eqnarray}
-\Big(\overline{\frac{B^{\prime}_{i} B^{\prime}_{j}}{4 \pi} - \delta_{ij} \frac{{B^{\prime}}^2}{8 \pi}}\Big) = -\nu^{\mathcal M}_{ij} \, {\bar \rho} \, \overline{ s_{ij}} \, ,
\label{4}
\end{eqnarray}
where ${\bar s}_{ij}= \frac {\partial \bar v_i}{\partial x_j} 
+ \frac {\partial \bar v_j}{\partial x_i} 
- \frac{2}{3} \nabla \cdot \bar{\bf v} \delta_{ij}$ is 
the strain tensor. The turbulent viscosities are parameterized
through an $\alpha$ prescription as 
\begin{eqnarray}
\nu^{\mathcal R}_{ij} \sim \alpha^{\mathcal R}_{ij} \, {\bar c_s} \, h , \, \, \, \, \,  \nu^{\mathcal M}_{ij} \sim \alpha^{\mathcal M}_{ij} \, {\bar c_s} \, h \, , 
\label{5}
\end{eqnarray}
where $\nu_{ij} = \nu^{\mathcal R}_{ij} + \nu^{\mathcal M}_{ij}$ is the net 
turbulent viscosity and $\alpha_{ij} = \alpha^{\mathcal R}_{ij} 
+ \alpha^{\mathcal M}_{ij}$ is the net turbulent viscosity parameter. With 
these parameterizations, the coupled accretion-outflow 
dynamical equations in quasi-stationary state are as follows:

(a) Mass transfer:
\begin{eqnarray}
\frac{\partial ({\bar \rho} \, {\bar v_j} )}{\partial x_j} = 0 \, .
\label{6}
\end{eqnarray}

We define the net mass flow rate which is a constant, through an 
integro-differential equation as 
\begin{eqnarray}
\int^{+h}_{-h} \int_{r} \int^{2 \pi}_{0} 
\Big[\frac{1}{r} \frac{\partial}{\partial r} (r {\bar \rho} \, {\bar v_r}) \,+ \, \frac{\partial}{\partial z} ({\bar \rho} \, {\bar v_z}) \Big]
 r d\varphi dr dz =  - {\dot M} \, ,
\label{7}
\end{eqnarray}

where the first term is the signature of the radial accretion flow and 
the second term attributes to outflow. If we discard 
${\bar v_z}$ (neglecting outflow), Eqn. (\ref{7}) reduces to 
a height integrated continuity equation of the accretion flow, and where ${\dot M}$  
would then be the usual mass accretion rate.

(b) Momentum transfer: 

The momentum balance equation in the mean field MHD is given by

\begin{eqnarray}
\frac{\partial \left( \bar \rho \bar v_i \bar v_j \right)}{\partial x_j}
\, = \, - \bar \rho \frac{\partial \varphi_G}{\partial x_i}
- \, \frac{\partial \bar p}{\partial x_i} 
+ \, \frac{\partial}{\partial x_j} \Big(\frac{\bar B_i \bar B_j}{4 \pi} 
- \delta_{ij} \frac{{\bar B}^2}{8 \pi} \Big) 
+ \, \frac{\partial}{\partial x_j} \Big(\overline{t^{\mathcal R}_{ij}}  
+ \overline{t^{\mathcal M}_{ij}} \Big) \, . 
\label{8}
\end{eqnarray}

Using Eqn. (\ref{6}) and integrating Eqn. (8) vertically, the radial momentum balance equation is given by 

\begin{eqnarray}  
\int^{+h}_{-h} \Bigg[\, \bar \rho \bar v_r \frac{\partial \bar v_r}{\partial r}
- \, \bar \rho \frac{{\bar \lambda}^{2}}{r^3}
+ \, \bar \rho \bar v_z \frac{\partial \bar v_r}{\partial z} 
+ \, \bar \rho F_{Gr} 
+ \, \frac{\partial \bar P}{\partial r} 
+ \frac{1}{4\pi} \Big(\frac{\bar {B_{\varphi}}}{r} \frac{\partial}{\partial r} (r \bar {B_{\varphi}})   
+ \, \bar {B_z} \frac{\partial \bar {B_z}}{\partial r} 
- \, \bar {B_z} \frac{\partial \bar {B_r}}{\partial z} \Big) \, \Bigg] \, dz \, 
= \, 0 \, .
\label{9}
\end{eqnarray}

In deriving this we have used divergence criteria of magnetic field. $F_{Gr}$ is the 
radial component of the gravitational force. In a similar fashion, we 
write the azimuthal momentum balance equation as 

\begin{eqnarray}
\int^{+h}_{-h} \Big(\,
\bar \rho \frac{\bar v_r}{r} \frac{\partial \bar \lambda}{\partial r} 
+ \, \bar \rho \frac{\bar v_z}{r} \frac{\partial \bar \lambda}{\partial z} \, \Big ) \, dz
= 
\, \frac{1}{r^2} \frac{\partial}{\partial r} \int^{+h}_{-h} \Big(r^2 {\bar t}_{r\varphi}\Big) \, dz 
+ \, \int^{+h}_{-h} \frac{1}{r^2} \frac{\partial}{\partial r} \Big(\frac{r^2 \bar {B_r} \bar {B_\varphi}}{4 \pi} \Big) \, dz 
+ \, \frac{\bar {B_\varphi} \bar {B_z}}{4 \pi} \, \Big \vert^{+h}_{-h} \, .   
\label{10}
\end{eqnarray}

The last term in the right hand side of Eqn. (\ref{10}) is the magnetic torque that acts 
at the accretion-outflow surface, and which helps in transporting 
the angular momentum vertically outwards. This term is responsible for the 
mass loss in the wind. 
Next we derive the vertical momentum balance equation which is obtained from 
the $z$ component of Eqn. (\ref{8}).

\begin{eqnarray}
2 \, \int^{h}_{0}
\Bigg[\bar \rho \bar v_r \frac{\partial \bar v_z}{\partial r} 
+ \, \bar \rho \bar v_z \frac{\partial \bar v_z}{\partial z} 
+ \, \bar \rho F_{Gz}  
+ \, \frac{\partial \bar P}{\partial z}  
+ \, \frac{\partial}{\partial z} \Big(\frac{{\bar B}^{2}_{\varphi} + {\bar B}^{2}_{r}}{8 \pi} \Big)
- \, \frac{\bar B_r}{4 \pi} \frac{\partial \bar B_z}{\partial r} \Bigg] \, 
dz \,  = \, 0 \, , 
\label{11}
\end{eqnarray}

where $F_{Gz}$ is the vertical component of the gravitational 
force. Equation (\ref{11}) is integrated from $0$ to $h$ due to the 
reflection symmetry of 
all the dynamical variables across the accretion-outflow coupled surface. 
Equation (\ref{11}) contains the information of the outflow dynamics within 
the accretion-outflow region. The matter starts to accelerate vertically 
outwards from just above the equatorial plane of the accretion region. 
If there is no outflow, then $v_z = 0$, and if we neglect the magnetic 
pressure and magnetic stresses, Eqn. (\ref{11}) reduces to the well 
known hydrostatic equilibrium 
condition in the disk, and the usual hydrostatic disk scale-height can be obtained. 

(c) Divergence condition:

\begin{eqnarray}
\int^{+h}_{-h} \Big[\frac{1}{r} \frac{\partial}{\partial r} (r {\bar B_r}) 
+ \, \frac{\partial \bar B_z}{\partial z} \Big] \,dz \, = \, 0 \, .
\label{12}
\end{eqnarray}

The divergence condition determines the symmetry 
property of magnetic field components. Whether the radial and the vertical 
component of the magnetic field will follow odd and even symmetry 
or vice-versa in the 
$z$ direction, can be ascertained from the above equation.  

(d) Magnetic induction:

The turbulent magnetic induction equation is derived from the
mean field MHD theory (e.g., [31]). Following 
the usual procedure and neglecting the dynamo effect, the steady state 
induction equation in the tensorial form is given by

\begin{eqnarray}
\epsilon_{ijk} \epsilon_{kmn} \frac{\partial}{\partial x_j}({\bar v_m} {\bar B_n}) \, - \, \epsilon_{ijk} \frac{\partial}{\partial x_j} 
\Big(\epsilon_{kmn} \nu^{\mathcal R}_{ml} \frac{\partial {\bar B_n}}{\partial x_l}\Big) \, = \, 0 \, , 
\label{13}
\end{eqnarray}

where $\nu^{\mathcal R}_{ml}$ is the kinetic part of the turbulent viscosity. The 
above equation has been written in the most general form 
considering an anisotropic turbulence. Splitting Eqn. (\ref{13}) in radial and 
azimuthal direction we obtain after vertical integration 

\begin{align} 
\int^{+h}_{-h} \Big[
\frac{\partial}{\partial z} ({\bar v_r} {\bar B_z} - {\bar v_z} {\bar B_r}) \, + \, \frac{\partial}{\partial z} \Big(\nu^{\mathcal R}_{zz} \frac{\partial {\bar B_r}}{\partial z} - \nu^{\mathcal R}_{rr} \frac{\partial {\bar B_z}}{\partial r}\Big) 
\Big ] \, dz \, = \, 0 \, . 
\label{14}
\end{align}
and 
\begin{eqnarray} 
\int^{+h}_{-h} \Big[
\frac{\partial}{\partial z} ({\bar v_{\varphi}} {\bar B_z} 
- {\bar v_z} {\bar B_{\varphi}}) 
- \, \frac{\partial}{\partial r} ({\bar v_r} {\bar B_{\varphi}} 
- {\bar v_{\varphi}} {\bar B_r}) 
+ \, \frac{\partial}{\partial z} \Big(\nu^{\mathcal R}_{zz} \frac{\partial {\bar B_{\varphi}}}{\partial z}\biggr) 
+ \, \frac{\partial}{\partial r} \biggl(\nu^{\mathcal R}_{rr} \frac{\partial {\bar B_{\varphi}}}{\partial r} 
+ \nu^{\mathcal R}_{\varphi \varphi} \frac{\bar B_{\varphi}}{r}\Big) \Big ] \, dz
 \, = \, 0 \, .
\label{15}
\end{eqnarray}

We do not show here the vertical component of the induction equation as it is 
similar to that of the radial equation, and contains the same 
information regarding 
the magnetic dynamics of the flow. Note that the turbulent diffusion in 
the induction equation arises only from the kinetic part of the turbulent 
stress tensor through the Reynolds stress. This is attributed to the 
mean field approximation, where we split the quantities in 
mean and turbulent part. 

(e) Energy conservation:

The Poynting flux $S_j$ is given by 

\begin{eqnarray}
S_j = \frac{c}{4 \pi} \epsilon_{jlm} E_l B_m\, .
\label{16}
\end{eqnarray}

Using mean field MHD and the induction equation, and by discarding the microscopic 
resistivity,  we write 

\begin{eqnarray}
\frac{\partial {\bar S}_j}{\partial x_j} 
= \, \frac{\bar v_j}{c} \epsilon_{jlm} {\bar B_l} {\bar J_m} 
+ \, \frac{\bar J_j}{c} \epsilon_{jlm} \overline{v^{\prime}_l B^{\prime}_m}
+ \, \frac{\bar v_j}{c} \epsilon_{jlm} \overline{B^{\prime}_l J^{\prime}_m} 
+ \, \frac{\bar B_j}{c} \epsilon_{jlm} \overline{J^{\prime}_l v^{\prime}_m} \, , 
\label{17}
\end{eqnarray}

where $J$ is the current density. In contrast to the other correlation terms, the last term in the right hand side of Eqn. (17) is not a first order correlation 
term, rather a second order or higher order correlation term [32]; we omit this term in our present study. 
Neglecting the kinetic dynamo effect, using Eqn. (17), the steady state 
energy conservation equation of a turbulent magnetic fluid in a 
tensorial form using mean field MHD is then given by 

\begin{align}
\frac{\partial}{\partial x_j} \Biggl[{\bar \rho} {\bar v_j} \biggl(\frac{{\bar v}^2}{2} + \frac{8 - 3 \beta}{2 \beta} \frac{\bar P_g}{\bar \rho} + \varphi_G \biggr) 
\, - \, {\bar v_i} {\overline{t^{\mathcal R}_{ij}}}  \Biggr]
\, + \frac{\bar v_j}{c} \epsilon_{jlm} {\bar B_l} {\bar J_m}  
\, + \frac{\bar v_j}{c} \epsilon_{jlm} \overline{B^{\prime}_l J^{\prime}_m}
- \, \frac{1}{4 \pi} \epsilon_{jlm} \nu^{\mathcal R}_{lk} \frac{\partial {\bar B_m}}{\partial x_k} \epsilon_{jlm} \frac{\partial {\bar B_m}}{\partial x_l} \, 
\, + \frac{\partial F^{r}_j}{\partial x_j} \, = \, 0 \, ,  
\label{18}
\end{align}

where $\beta = {\bar P_g}/({\bar P_g} + {\bar P_r})$, the ratio of gas pressure 
to the total pressure in the accretion-outflow coupled region is assumed to be 
constant. ${\bar P_r}$ is the radiation pressure in the gas-radiation mixture. 
The radiation field has been assumed to be locally isotropic. The ratio of specific 
heat of the fully ionized gas and radiation are taken as $\gamma_g = 5/3$ 
and $\gamma_r = 4/3$ respectively. The `effective ratio of specific 
heat' $\Gamma$ is related to $\beta$ through   
$\Gamma = ({8 - 3 \beta})/({6 - 3 \beta})$. In deriving Eqn. (\ref{18}), we have 
neglected the turbulent thermal conductivity. The last term in the above equation 
represents the transport of radiative flux. Multiplying Eqn. (\ref{8}) by ${\bar v_j}$ and using the continuity equation, 
the vertically integrated energy budget for accretion-induced outflow 
is thus obtained below

\begin{align}
\int^{+h}_{-h} \Big[
\frac{3}{2} (2 - \beta){\bar \rho} \, {\bar v_r} \frac{\partial {\bar c}^2_s}{\partial r} 
\, -  {\bar v_r} {\bar c}^2_s \frac{\partial {\bar \rho}}{\partial r}  \Big] dz 
\, 
+ \int^{+h}_{-h} \Big[ \frac{3}{2} (2 - \beta){\bar \rho} \, {\bar v_z} \frac{\partial {\bar c}^2_s}{\partial z} 
\, -  {\bar v_z} {\bar c}^2_s \frac{\partial {\bar \rho}}{\partial z} \Big] \, dz
\, \nonumber \\
= \, \int^{+h}_{-h}  \Big[ {\bar \rho} \, \nu^{\mathcal R}_{r\varphi} 
\, \bar{s}^{2}_{r\varphi} 
\, +   \frac{1}{4 \pi} \epsilon_{jlm} \nu^{\mathcal R}_{lk} \frac{\partial {\bar B_m}}{\partial x_k} \epsilon_{jlm} \frac{\partial {\bar B_m}}{\partial x_l}   
\Big] \, dz   - 2 F^{r+} \, .
\label{19}
\end{align}

Left hand side of the above equation is the signature of advection of energy 
flux in radial and vertical direction due to accretion and outflow respectively. 
First and the second term of the right hand side expresses the turbulent 
viscous heating due to $r \varphi$ component of the stress tensor 
and the turbulent Ohmic dissipation or Joule's heat loss 
respectively in the accretion flow. Turbulent Ohmic dissipation symbolizes 
a resistive flow due to which constant annihilation of the magnetic flux occurs. 
The last term of the equation is the 
flux of radiation that is escaping from the accretion-outflow surface.  
Similar to that of the induction Eqn. (\ref{13}), the turbulent diffusion in 
the energy equation also arises only through the Reynolds stress tensor

The net heat flux generated in the accretion flow is defined by 

\begin{eqnarray}
q^{+} \, = \, \int^{+h}_{-h}  \bigg( {\bar \rho} \, \nu^{\mathcal R}_{r\varphi} 
\, \bar{s}^{2}_{r\varphi} 
\, +   \frac{1}{4 \pi} \epsilon_{jlm} \nu^{\mathcal R}_{lk} \frac{\partial {\bar B_m}}{\partial x_k} \epsilon_{jlm} \frac{\partial {\bar B_m}}{\partial x_l}   
\bigg) \, dz \, .
\label{20}
\end{eqnarray}

Defining $f$ as a constant cooling factor scaled through a relation  

\begin{eqnarray}
q^{+} - 2 F^{r+} \, = \, (1-f) q^{+}  \, ,
\label{21}
\end{eqnarray}

the final form of the heat flux equation is obtained by coupling Eqns. (\ref{19}),   
(\ref{20}) and (\ref{21}). A constant $f$ implies that the radiative 
heat loss from the accretion surface has a linear proportionality with the 
net heat flux generated in the system. Although $f$ should vary 
radially, however, within a small inner accretion region where the 
accretion and outflow are coupled, 
this presumption is an acceptable approximation. The parameter $f$ 
reflects the extent to which 
the accretion flow is advective. $f \rightarrow 1$ represents an 
accretion flow with efficient cooling or is radiatively efficient. At the other 
extremity, $f \rightarrow 0$ epitomize a strongly advective or advection dominated system which are 
radiatively inefficient. 

Equations (\ref{7}), (\ref{9}), (\ref{10}), (\ref{11}), (\ref{12}), (\ref{14}), 
(\ref{15}) and (\ref{19}) exhibit the dynamical behavior of 
accretion-induced outflow in a resistive MHD paradigm in mean field approximation. 
In the next section, we explore a technique to solve these eight coupled partial 
differential equations using Eqns. (\ref{5}), (\ref{20}) and (\ref{21}), rigorously.

\section{Polynomial expansion and solution procedure}

The equations describing the turbulent magneto-fluid in the last section 
are extremely complicated and it is beyond our scope to solve them numerically. 
The only procedure left in our hand is to explore some kind of approximate 
analytical method or quasi-analytical method to solve them. 
We follow a power law self-similar approach (see also NY94) with a 
polynomial expansion to solve the equations to obtain the class 
of solutions. For the present
purpose we invoke a generalized n$^{\textnormal{th}}$ 
degree polynomial expansion of all the dynamical 
variables, where the flow variables are functions of both radial and 
vertical coordinate. We restrict ourselves in Newtonian paradigm as power law 
self-similar 
solutions are valid only in the limit of Newtonian approximation. The generalized 
Newtonian potential at any ($r,z$) in the accretion-outflow region is written in 
the form of power series  
\begin{eqnarray}
\varphi_G \, (r,z)  \, =  \, -GM(r^{-1} - \frac{1}{2} r^{-3} z^2 
+ \frac{3}{8} r^{-5} z^4 - \cdots) \, , 
\label{22}
\end{eqnarray}
where $M$ is the mass of the central object. The gravitational force in the radial and vertical direction is then 
written in the form of a polynomial expansion as shown below.
\begin{eqnarray}
F_{Gr}(r, z) = G M \sum_{n=0}^{\infty} {-3/2 \choose n} r^{-2 - 2n} z^{2n} \, .
\label{23}
\end{eqnarray}
\begin{eqnarray}
F_{Gz} (r, z) = G M \sum_{n=0}^{\infty} {-3/2 \choose n} r^{-3 - 2n} z^{2n + 1} \, .
\label{24}
\end{eqnarray}

$F_{Gr}$ and $F_{Gz}$ have even and odd symmetry in $z$ direction, 
respectively. We seek a polynomial expansion in the similar form for 
all dynamical variables, where the flow of matter in the accretion region is being considered 
to have reflection symmetry about the equatorial plane.  All the hydrodynamical variables 
will have even symmetry except those 
directly related to $F_{Gz}$. The outflow velocity $v_z$ will then have a odd 
symmetry in $z$ direction. Consequently, the radial component of the magnetic field $\bar B_r$ and the vertical component 
of the magnetic field $B_z$ will have even and odd symmetry configuration, respectively, 
which is required from the divergence 
condition of the magnetic field. Odd symmetry configuration of magnetic 
field has been used previously on other occasions in context to outflows/jets from the accretion 
flows (e.g., [33,26]). The flow velocities, angular momentum and 
the density are then written in the following polynomial form 
\begin{align} 
\bar v_{r} (r, z) = \sum_{n=0}^{\infty} v_{r2n} r^{a - 2n} z^{2n} &,   
\bar v_{\varphi} (r, z) = \sum_{n=0}^{\infty}  v_{\varphi 2n} r^{b - 2n} z^{2n}, \nonumber \\ 
\bar {\lambda} (r, z) = \sum_{n=0}^{\infty}  v_{\varphi 2n} r^{b - 2n+1} z^{2n} &,  
\bar v_{z} (r, z) = \displaystyle\sum_{n=0}^{\infty} v_{z(2n+1)} r^{c - 2n} z^{2n+1},  \nonumber \\
\bar c_{s} (r, z) = \sum_{n=0}^{\infty}  c_{s2n} r^{d - 2n} z^{2n} &,  
\bar {\rho} (r, z) = \sum_{n=0}^{\infty}  \rho_{2n} r^{e - 2n} z^{2n} .
\label{25}
\end{align}
Similarly, the components of the magnetic field can be expanded as 
\begin{align}
\bar B_{r} (r, z) = \sum_{n=0}^{\infty}  B_{r2n} r^{i - 2n} z^{2n}, \, \, 
\bar B_{\varphi} (r, z) = \sum_{n=0}^{\infty} B_{\varphi 2n} r^{j - 2n} z^{2n}, 
\bar B_{z} (r, z) = \sum_{n=0}^{\infty} B_{z(2n +1)} r^{k - 2n} z^{2n+1}\, ,
\label{26}
\end{align}
where, $v_{r2n}, \, v_{\varphi 2n},  \, v_{z(2n+1)},  \, c_{s2n}, \, \rho_{2n},  
\, B_{r2n},   \, B_{\varphi 2n},  \, B_{z(2n +1)}$ are the dimensionless 
coefficients which will be evaluated from the MHD conservation 
equations. 

We determine the exponents $a, b, c, d, e, i, j, k$ by self 
comparison of various terms in the model equations. Substituting 
the solutions from Eqns. (\ref{25}) and (\ref{26}) in the MHD conservation 
equations and comparing the exponents of $r$ and $z$, we 
obtain $ a = -1/2, \, b = -1/2, \, c = -3/2, \, 
d = -1/2, \, e = -3/2, \, i = -5/4, \, j = -5/4, \, k = -9/4$.

Using the above value of exponents and using the polynomials 
in Eqns. (\ref{25}) and (\ref{26}), Eqns. (\ref{9}), (\ref{10}), 
(\ref{11}), (\ref{12}), (\ref{14}), (\ref{15}) and (\ref{19}) can be 
expanded in the power of aspect ratio $(h/r)$. Using a simple technique given by  
the form
\begin{align}
\sum_{n=0}^{\infty} u_n x^n \, \sum_{n=0}^{\infty} v_n x^n  
\, \sum_{n=0}^{\infty} w_n x^n 
= \, \sum_{n=0}^{\infty} x^n \, \sum_{m=0}^{\infty} \, 
\sum_{l=0}^{\infty} \, u_{n-m} \, v_{m-l} \, w_l \, , 
\label{27}
\end{align}
the above polynomial 
equations then can be written in a generic form as 
\begin{eqnarray}
A_0 \, \Big(\frac{h}{r}\Big)^0 + A_1 \, \Big(\frac{h}{r}\Big)^2 
+ A_2 \, \Big(\frac{h}{r}\Big)^4 + \cdots \, = \, 0 \, ,
\label{28}
\end{eqnarray} 
where, $A_0, A_1, A_2 , \cdots$ are zeroth order, first order, second order 
and higher order coefficients which are nonlinear functions of 
$v_{r2n}, \, v_{\varphi 2n}, \, v_{z(2n+1)}, \, c_{s2n}, \, \rho_{2n}, 
\, B_{r2n}, \, B_{\varphi 2n},  \, B_{z(2n +1)}$ corresponding to $n=0, 
\, n = 1, \, n = 2 , \cdots$, respectively. Equation (\ref{28}) is a 
linear combination in powers of 
$h/r$ which are linearly independent. 

If we neglect all the terms of the order $\geq (h/r)^2$, and only keep the 
zeroth order term, the generic Eqn. (\ref{28}) after neglecting the magnetic 
field contribution will reduce to algebraic equations in NY94. To exemplify, 
in the appendix, we have shown the polynomial expansion of the 
integro-differential mass transfer Eqn. (\ref{7}) and the radial momentum 
balance Eqn. (\ref{9}) explicitly. Even if the accretion 
flow has considerable thickness, $h$ in general would always be less 
than $r$. Further, magnetic stresses will compress or squeeze the accretion region. 
Considering the expression in Eqn. (\ref{28}) up to the term $(h/r)^2$ would then be a reasonable approximation. 
Restricting the expansion up to $(h/r)^2$ corresponding to $n=1$ and 
neglecting the terms with orders $\geq \, (h/r)^4 $ in the generic expression (\ref{28}), 
we equate $A_0$ and $A_1$ to zero, respectively. Extending this to MHD conservation equations and 
assuming isotropic turbulence, after rigorous algebra,  
we will then have fifteen independent non-linear algebraic equations 
with sixteen unknown coefficients consisting of zeroth and first order only, which are shown below. \\ \\
Equation (\ref{7}) renders
\begin{align}
\rho_0 (v_{r0} -v_{z1}) \,t  + \left[\rho_0 (v_{r2} -v_{z3})  + \rho_2 (v_{r0} -v_{z1}) \right] \frac{t^3}{3} \, = \, - \frac{\dot M}{4\pi} \, .
\label{29}
\end{align}
Equation (\ref{12}) renders
\begin{eqnarray}
B_{z1} = \frac{B_{r0}}{4}\, ,
\label{30}
\end{eqnarray}  
\begin{eqnarray}
B_{z3} = \frac{3}{4} {B_{r2}}\, .
\label{31}
\end{eqnarray} 
Equation (\ref{9})  renders
\begin{eqnarray}
-\frac{1}{2} v^{2}_{r0} - v^{2}_{{\varphi}0} + GM -  \frac{5}{2} c^{2}_{s0} 
- \frac{1}{16\pi} \frac{B^{2}_{{\varphi}0}}{\rho_0} \, = \, 0 \, ,
\label{32}
\end{eqnarray}
\begin{align}
-\frac{1}{2} {\rho_2} \, v^{2}_{r0} - 3 {\rho_0} \, v_{r0} \, v_{r2} 
- {\rho_2} \, v^{2}_{{\varphi}0} - 2 {\rho_0} \, v_{\varphi 0} \, v_{\varphi 2}
+ 2 {\rho_0} \, v_{z1} \, v_{r2} 
+ GM(\rho_2 - \frac{3}{2} \rho_0) 
- \frac{9}{2} ({\rho_2} \, c^{2}_{s0} + 2 {\rho_0} \, c_{s0} \, c_{s2}) \nonumber \\
+ \frac{1}{4\pi} (- \frac{5}{2} B_{\varphi 0} \, B_{\varphi 2} 
- \frac{9}{4} B^{2}_{z1}
- 2 B_{z1} \, B_{r2} ) \, = \, 0 \, .
\label{33}
\end{align}
Equation (\ref{10}) renders
\begin{eqnarray}
\frac{1}{2} v_{r0} \,v_{{\varphi}0} + \frac{3}{4} \alpha_{r\varphi} 
\, t \, v_{{\varphi}0} \, c_{s0} 
+ \frac{1}{16\pi} \frac{B_{r0} \,B_{{\varphi}0}}{\rho_0} \, = \, 0 \, ,
\label{34}
\end{eqnarray}
\begin{eqnarray}
\frac{1}{2} ({\rho_2} \, v_{r0} \, v_{\varphi 0} 
+ {\rho_0} \, v_{\varphi 0} \, v_{r2}) - \frac{3}{2} {\rho_0} \, v_{r0} \, v_{\varphi 2} 
+ 2 {\rho_0} \, v_{z1} \, v_{\varphi 2} 
+ \frac{1}{2} \alpha_{r\varphi} \, t \,
\left(\frac{3}{2} {\rho_2} \, c_{s0} \, v_{\varphi 0}  
+   \frac{3}{2} {\rho_0} \, c_{s2} \, v_{\varphi 0}  +  \frac{7}{2} {\rho_0} \, c_{s0} \, v_{\varphi 2} \right) \nonumber \\
+ \frac{1}{16\pi} (B_{{\varphi}0} \, B_{r2} + 7 B_{r0} \, B_{{\varphi}0}) \, = \, 0 \, .
\label{35}
\end{eqnarray} 
Equation (\ref{11}) renders
\begin{eqnarray}
- \frac{3}{2} {\rho_0} \, v_{r0} \, v_{z1} + {\rho_0} \, v^{2}_{z1} + {\rho_0} \,GM
+ 2(\rho_2 \, c^{2}_{s0} + 2 \rho_2 \, c_{s0} \, c_{s2}) 
+ \, \frac{1}{4\pi} (2 B_{r0} \, B_{r2} + 2 B_{\varphi 0} \, B_{{\varphi}2} 
+ \frac{9}{4} B_{r0} \, B_{z1} )  \, = \, 0 \, ,
\label{36}
\end{eqnarray}
\begin{align}
- \frac{3}{2} {\rho_2} \, v_{r0} \, v_{z1} - \frac{3}{2} {\rho_0} \, v_{r2} \, v_{z1} - \frac{7}{2} {\rho_0} \, v_{r0} \, v_{z3} + {\rho_2} \, v^2_{z1} 
+ 4 {\rho_0} \, v_{z1} \, v_{z3} 
+  \, GM(\rho_2 - \frac{3}{2} \rho_0) 
+ 4 ({\rho_2} \, c_{s0} \, c_{s2} + {\rho_0} \, c^2_{s2}) \nonumber \\
+ \frac{1}{8\pi} (4 B^2_{r2} + 4 B^2_{\varphi 2})  
- \frac{1}{4\pi} \left(-\frac{9}{4} B_{r2} \, B_{z1} - \frac{17}{4} B_{r0} \, B_{z3} \right) \, = \, 0 \, .
\label{37}
\end{align}
Equation (\ref{14}) renders
\begin{eqnarray}
(v_{r0} \,  B_{z1} - v_{z1} \, B_{r0}) + \alpha^{\mathcal R}_{r\varphi} \, 
t \left(2 c_{s0}\,B_{r2} + \frac{9}{4} c_{s0} \,B_{z1} \right) \, = \, 0 \, .
\label{38}
\end{eqnarray}
\begin{eqnarray}
(v_{r2} \,  B_{z1} + v_{r0} \, B_{z3}) - (v_{z3} \, B_{r0} + v_{z1} \, B_{r2}) 
+ \, \alpha^{\mathcal R}_{r\varphi} \, t 
\left(2 c_{s2} \,B_{r2} + \frac{9}{4} c_{s2} \,B_{z1}  
+ \frac{17}{4} c_{s0} \,B_{z3} \right) \, = \, 0 \, .
\label{39}
\end{eqnarray}
Equation (\ref{15}) renders
\begin{eqnarray}
(v_{\varphi 0} \, B_{z1} - v_{z1} \, B_{\varphi 0}) 
+ \frac{7}{4} (v_{r0} \, B_{\varphi 0} - v_{\varphi 0} \, B_{r0}) 
+ \alpha^{\mathcal R}_{r\varphi} \, t 
\left(2 c_{s0} \,B_{\varphi 2} + \frac{7}{16} c_{s0} \,B_{\varphi 0} \right) \, = \, 0 \, ,
\label{40}
\end{eqnarray}
\begin{eqnarray}
3(v_{\varphi 0} \, B_{z3} - v_{\varphi 2} \, B_{z1})  
- 3(v_{z3} \, B_{\varphi 0} - v_{z1} \, B_{\varphi 2})
+ \, \frac{7}{4} \left[(v_{r0} \, B_{\varphi 2} + v_{r2} \, B_{\varphi 0}) 
- \, (v_{\varphi 2} \, B_{r0} + v_{\varphi 0} \, B_{r2}) \right] \nonumber \\
+ \alpha^{\mathcal R}_{r\varphi} \, t \left[6 \,c_{s2} \,B_{\varphi 2} 
+ \frac{15}{16} (c_{s2} \,B_{\varphi 0} + 9 \, c_{s0} \,B_{\varphi 2}) \right]\, = \, 0 \, .
\label{41}
\end{eqnarray}
Equation (\ref{19}) using Eqns. (\ref{20}) and (\ref{21}) renders
\begin{eqnarray}
\frac{3}{2} (\beta -1) \, v_{r0} \, c_{s0} \,
= \, (1-f) \,\alpha^{\mathcal R}_{r\varphi} \, t \, \Big(\frac{9}{4} v^2_{\varphi 0}
+  \frac{1}{16} \frac{B^2_{\varphi 0}}{4\pi \rho_0} \Big) \, ,
\label{42}
\end{eqnarray}
\begin{align}
\frac{3}{2} (\beta -1) \, \rho_0 \, v_{r2} \, c^2_{s0} 
+ (9 \beta -15) \, \rho_0 \, v_{r0} \, c_{s0} \, c_{s2}  
+  \frac{1 + 3 \beta}{2} \, \rho_2 \, v_{r0} \, c^2_{s0} 
+  \, 6 (2 - \beta) \, \rho_0 \, v_{z1} \, c_{s0} \, c_{s2} 
- \, 2 \rho_2 \, v_{z1} \, c^2_{s0} \nonumber \\
= \, (1-f) \, \alpha^{\mathcal R}_{r\varphi} \, t \, 
\Bigg[ \Big( \frac{9}{4} \rho_2 \, c_{s0} \, v^2_{\varphi 0} 
+ \,  \frac{9}{4} \rho_0 \, c_{s2} \, v^2_{\varphi 0} 
+ \frac{21}{2} \rho_0 \, c_{s0} \, v_{\varphi 0} \, v_{\varphi 2} \Big) 
+ \, \frac{1}{4\pi}  \,
\Big (4 \, c_{s0} \, B^2_{\varphi 2} + 4 \, c_{s0} \, B^2_{r2} 
+ \frac{81}{16} c_{s0} \, B^2_{z1} + 9 \,c_{s0} \, B_{r2} \, B_{z1} \nonumber \\
+  \frac{1}{16} c_{s2} \, B^2_{\varphi 0} 
+  \frac{9}{8}c_{s0} \, B_{\varphi 0} \, B_{\varphi 2} \Big) \Bigg] \, .
\label{43}
\end{align}

In the above equations $t = h(r)/r$. Using Eqns. (\ref{29}), (\ref{30}), 
(\ref{31}), (\ref{33}), (\ref{34}), (\ref{35}), (\ref{36}), 
(\ref{38}), (\ref{39}), (\ref{40}), (\ref{42}), and after very complicated 
and tedious algebra, we systematically determine the value of all the first order 
coefficients in terms of zeroth order coefficients of 
hydrodynamic variables. Substituting them in Eqns. (\ref{32}), (\ref{37}), 
(\ref{41}) and (\ref{43}), we successfully reduce fifteen equations to 
four nonlinear algebraic equations comprising of zeroth order 
coefficients $v_{z1}, v_{\varphi 0}, B_{r0}$ and $B_{\varphi 0}$ only. We then 
solve the above four nonlinear equations through an iterative Newton 
Raphson method for appropriate initial guess. In these way we 
can obtain the values of zeroth and first order coefficients 
of the corresponding dynamical variables. The value of the 
coefficients are scaled 
by putting $G=M=1$, and $\dot M$ in units of $\dot M_{\rm Edd}$.

In the next section, we evaluate them for appropriate choice 
of parameters $\dot M, \, \beta$ and $f$. 

\section{Evaluation of the coefficients}

In \S \RNum{1} we have analyzed the necessity 
to have a sub-Keplerian advective accretion regime 
to eject strong outflows and jets from the accretion region in the vicinity 
of the BHs. The corresponding accretion regime 
can be possibly envisaged if the mass accretion rate or net mass 
flow rate is considerably sub-Eddington ($\dot M \, \lsim \, 10^{-2} \dot M_{\rm Edd}$). The 
ratio of gas to the total pressure $\beta$ and the radiative cooling factor 
$f$ directly depends on $\dot M$. Looking meticulously into the equations 
from ($29-43$), we notice that the continuity  equation is written 
in the integral form through Eqn. (\ref{29}), unlike the other 
hydrodynamical equations. This is being 
deliberately done to preserve the information of $\dot M$ in the flow, 
as $\dot M$ is the most fundamental parameter which determines the nature of the 
BH accretion paradigm. However, this constraints the number of 
dynamical equations. For our case, we have sixteen unknown coefficients but 
fifteen number of equations. To resolve this, we proceed in the following way. 

$\dot M$ carries the information of the density of the flow which means that  
if $\dot M$ is known, in principle, density too is known. 
Equation (\ref{29}) reveals that the density in the accretion-outflow 
region is a function of two unknown coefficients $\rho_0$ and $\rho_2$. 
$\rho_0$ is the signature of inflow whereas $\rho_2$ is that of the outflow. If 
the outflow is discarded, $\rho_2$ looses its significance 
and $\dot M$ then becomes the usual mass accretion rate from 
where $\rho_0$ can be easily calculated. If 
$r_j$ represents the outer radial boundary of the accretion-outflow coupled 
region, then at 
$r \, \geq \, r_j$ the net mass flow $\dot M$ is equivalent 
to the mass accretion rate of the flow. At $r \, \geq \, r_j$, density of 
the accretion flow is just a function of  
$\rho_0$, which is then given by (also see [5])
\begin{eqnarray}
\rho_0 = \frac{\Big[5 + 2 \, \frac{1-\beta}{1-f} \, 
\frac{\alpha_{r\varphi}}{\alpha^{\mathcal R}_{r\varphi}} \Big ]^{3/2}}{12 \pi \sqrt{2}}
\frac{\dot M}{\alpha_{r\varphi}} \, .
\label{44}
\end{eqnarray}
$\alpha^{\mathcal R}_{r\varphi}$ arises due to mean field approximation. 
Note that, the value of $\rho_0$ computed in  Eqn. (\ref{44}) is not 
exactly the same as that of $\rho_0$ in Eqn. (\ref{29}), where it is coupled 
to $\rho_2$. However, we 
presume that the density in the accretion flow does not change abruptly due to 
emanation of the outflow and jet, as only a small fraction of matter is ejected out through the outflow. With this 
presumption, we calculate $\rho_0$ in Eqn. (\ref{29})
using Eqn. (\ref{44}), and supply its value to the rest of other 
nonlinear equations for further computation.   

In the usual accretion flow, when there is no net vertical flux, the vertical 
height of the accretion geometry is calculated from the hydrostatic 
pressure balance, assuming 
the pressure and density at the outer accretion surface to be zero. However, this 
physical condition cease to exist when the outflow is incorporated 
in the system, and the scale-height 
of the coupled accretion-induced outflow region becomes difficult to 
ascertain. In these 
circumstances accretion-outflow coupled surface can be treated as a photospheric
height, delineating between the accretion-outflow surface and the transition 
region leading to the outflow decoupling from the inflow. We treat 
$t = h(r)/r$ as a parameter to get physical plausible solutions. 

Pessah, Chan and Psaltis [34] had shown that during the exponential growth of 
magnetorotational instability at saturation, the ratio of Maxwell stress 
to Reynolds stress becomes 
\begin{eqnarray}
\frac{\overline{t^{\mathcal M}_{r\varphi}}}{\overline{t^{\mathcal R}_{r\varphi}}} \, = 
\, \frac{4-q}{q} \, , 
\label{45}
\end{eqnarray}
where $ 3/2 \, \leq \, q < 2 $. $q$ is related to the angular 
velocity through the relation $\Omega \, \sim \, r^{-q}$. $q \, = \, 3/2$
signifies an Keplerian flow whereas $q  > 3/2$ implies a sub-Keplerian 
or an advective accretion flow. Thus for a strongly advective flow 
$q \, >> \, 3/2$. We use the relation given in Eqn. (\ref{45}) 
to determine the value of $\alpha^{\mathcal R}_{r\varphi}$ used in our 
equations, which is then related to 
$\alpha_{r\varphi}$ as 
\begin{eqnarray}
\frac{\alpha^{\mathcal R}_{r\varphi}}{\alpha_{r\varphi}} \, = \, \frac{q}{4} \, .
\label{46}
\end{eqnarray}

The qualitative feature of the dynamical coefficients 
for an accretion-induced  outflow which are physically 
plausible should satisfy the following properties of the flow 
variables; $ \bar v_r \rightarrow$ negative, $\bar v_z \rightarrow$ 
positive, $\bar B_{\varphi 0} \rightarrow$ negative. All the other variables 
should have a positive value. The positivity of $\bar B_r$ and $\bar B_z$ 
is related to 
the open magnetic field lines threading the accretion flow, across which 
the accreting matter diffuses, and then gets accelerated outwards along 
the poloidal field lines by extracting the angular momentum from the 
flow. The term $-r^2 (\bar B_{\varphi} \bar B_z )_h$ [last term in Eqn. (\ref{10})], 
which is a magnetic torque on the accretion flow is attributed to  
the transport of angular momentum from the $±h(r)$ surfaces of the 
accretion-outflow region outwards. This 
term should be positive in order to launch a jet, which removes angular 
momentum from the flow and decouples from the accretion region. This 
premise makes a 
obvious choice for $\bar B_{\varphi}$ to have a negative value. It should 
be reminded that the negative value of many quantities do not necessarily 
mean that their magnitude is negative, but it represents the direction of their  
flow. 

Next we compute the value of the coefficients 
of all dynamical variables for three relevant choices of $\dot M$ with appropriate  
values of $\beta$ and $f$, conducive to form outflows and jets.

\subsection{Case 1. For $\dot M = 10^{-4} \, \dot M_{\rm Edd} $ }

The particular choice of $\dot M$ corresponds to RIAF, which is  
linked observationally to low-hard state of BHXRBs and LERGs/LLAGNs. This 
type of flow is significantly gas pressure dominated, and strongly advective. The 
flow is considerably geometrically thick, optically thin and 
radiatively inefficient. We choose 
appropriate values of $\beta$ and $f$ corresponding to this $\dot M$ to  
get physical plausible solutions. As this flow is highly sub-Keplerian
and strongly advective, an appropriate choice of value of $q \sim 1.9$ has been taken. In tables 1.1, 1.2 and 
1.3, we present the computed values of the dynamical coefficients 
for $\beta \, = \, 0.95$ , $f \,= \, 0.1$ with suitable 
choices of $\alpha$ and $t$. 
For $\beta \, \sim \, 0.95$, $\Gamma$ is $\sim 1.635$.
It is interestingly found that for $q \, << \, 1.9$ and in the range of $1.9 \, << \, q < 2$, 
we never found any physically valid solutions. However, for much stronger advective paradigm  
($\dot M \, << \, 10^{-4} \,  \dot M_{\rm Edd}$), one would still obtain physically correct solutions for $1.9 \, << \, q < 2$.   

We will analyze the family of analytical solutions later, however, we note that 
the values of the coefficients are in conformity with desired physically 
valid solutions (mentioned earlier). We notice that they  
are obtained only at high $\alpha$ ($\gsim \, 0.3$) and at a reduced 
vertical scale-height ($t \sim 0.1$) of the accretion-outflow coupled region. 
Although $t$ is small, it corresponds to a geometrically thick 
accretion flow ($t \, \gsim \, 0.1$, $t$ measures the degree of 
flow thickness). We have elucidated the necessity 
of a geometrically thick accretion flow to eject outflow and jet in previous 
paragraphs.  
However, corresponding to $\alpha \sim 0.3$, 
we never found any physically acceptable solution for 
$t > 0.1$. For higher $\alpha \sim 0.5$, however,  
we obtain a solution at a maximum value of $t \sim 0.2$. This infers that with 
the increase of $\alpha$, the plausible physical solutions can be obtained 
with a thicker accretion geometry. Nevertheless, very high $\alpha$ ($ > 0.5$) 
accretion flow might not be realistic in nature, and hence we restrict our 
study to a maximum plausible value of $\alpha \sim 0.5$. 

\begin{table*}
\large
\centerline{\large Table 1.1}
\centerline{\large $\dot M = 10^{-4} \dot M_{\rm Edd} $, $\beta = 0.95$, $f= 0.1$,  $\alpha = 0.3$, $t = 0.1$, $q=1.9$.}
\begin{center}
\begin{tabular}{cccccccc}
\hline
\hline
$\rho_0$  & $v_{r0}$  & $v_{\varphi 0}$ & $v_{z1}$  & $c_{s0}$  & $B_{r0}$ & $B_{\varphi 0}$ & $B_{z1}$ \\
\hline
7.4864e-5  &  -0.0899 & 0.3692  & 0.7069 & 0.5867 &  0.0036 & -0.0050 & 1.3e-3  \\
\hline
$\rho_2$  & $v_{r2}$  & $v_{\varphi 2}$ & $v_{z3}$  & $c_{s2}$  & $B_{r2}$ & $B_{\varphi 2}$ & $B_{z3}$ \\
\hline
-3.4889e-4  &  -0.2421 & 0.3471 & -33.1161 &  -0.3551 & 0.0523 & -0.2176 & 0.0793 \\ 
\hline
\hline
\end{tabular}
\end{center}
\end{table*}

\begin{table*}
\large
\centerline{\large Table 1.2}
\centerline{\large $\dot M = 10^{-4} \dot M_{\rm Edd} $, $\beta = 0.95$, $f= 0.1$,  $\alpha = 0.5$, $t = 0.1$, $q=1.9$.}
\begin{center}
\begin{tabular}{cccccccc}
\hline
\hline
$\rho_0$  & $v_{r0}$  & $v_{\varphi 0}$ & $v_{z1}$  & $c_{s0}$  & $B_{r0}$ & $B_{\varphi 0}$ & $B_{z1}$ \\
\hline
4.4918e-5  &  -0.1494 & 0.3694 & 0.3859 &  0.5891 & 0.0051 & -0.0041 & 4.25e-3  \\
\hline
$\rho_2$  & $v_{r2}$  & $v_{\varphi 2}$ & $v_{z3}$  & $c_{s2}$  & $B_{r2}$ & $B_{\varphi 2}$ & $B_{z3}$ \\
\hline
-1.1174e-4  &  -0.0925 & 0.3211  & -7.0486 & -0.2300 &  0.0257 & -0.0603 & 0.0193  
\\ 
\hline
\hline
\end{tabular}
\end{center}
\end{table*}

\begin{table*}
\large
\centerline{\large Table 1.3}
\centerline{\large \large $\dot M = 10^{-4} \dot M_{\rm Edd} $, $\beta = 0.95$, $f= 0.1$,  $\alpha = 0.5$, $t = 0.2$, $q=1.9$.}
\begin{center}
\begin{tabular}{cccccccc}
\hline
\hline
 $\rho_0$  & $v_{r0}$  & $v_{\varphi 0}$ & $v_{z1}$  & $c_{s0}$  & $B_{r0}$ & $B_{\varphi 0}$ & $B_{z1}$ \\
\hline
4.4918e-5  &  -0.2876 & 0.3636 & 0.6211 &  0.5977 & 0.0149 & -0.0061 & 5.2e-3  \\
\hline
$\rho_2$  & $v_{r2}$  & $v_{\varphi 2}$ & $v_{z3}$  & $c_{s2}$  & $B_{r2}$ & $B_{\varphi 2}$ & $B_{z3}$ \\
\hline
-2.1111e-4  &  -0.7225 & 0.6940  & -9.7460 & -0.3859 &  0.0598 & -0.0724 & 0.0448 \\
\hline
\hline
\end{tabular}
\end{center}
\end{table*}

Obtaining physically valid solutions of an accretion-induced outflow at a reduced 
geometrical thickness (scale-height) of the accretion region, as
compared to a non-magnetized accretion flow (without outflow) like ADAFs for a similar accretion paradigm, is 
owing to the fact, that the magnetic stresses in the flow has a tendency to squeeze or compress 
the region by acting oppositely to thermal pressure gradient, consequently reducing the scale-height of the 
accretion region. This squeezing effect has been discussed 
by several other authors (e.g., [25,30]) in context to magnetized 
accretion flow. This can be noticed from the vertical momentum balance equation (Eqn. \ref{11}). As we intend 
to see the effect of magnetic field on the geometrical thickness of the accretion region, for simplicity we ignore the 
outflow in (Eqn. \ref{11}). Further, we found that with the increase in $z$, the magnitude of all 
components of the magnetic field increases, however the density and the thermal pressure decreases with increase in 
$z$. As we ignore the outflow, for simplicity of our calculation, we consider density and thermal pressure at accretion flow scale-height ($h$) to be negligible as compared to their equatorial values; this will not alter the qualitative nature of our argument. Expanding the terms in (Eqn. \ref{11}), and restricting up to ${h^2}/{r^2}$, (Eqn. \ref{11}) will reduce to the 
magnetohydrostatic equilibrium equation, given by 
\begin{align}
{\bar \rho}_{\rm eq} \, \frac{h^2}{r^3} \sim {\bar P}_{\rm eq} \, - \frac{1}{8\pi} r^{-5/4} \,  
\left(B_{\varphi 0}  B_{\varphi 2} + B_{r0}  B_{r2} + \frac{9}{4} B_{r0}  B_{z1} \right) \, \frac{h^2}{r^2} \, , \nonumber \\
\label{46} 
\end{align}
where, ${\bar \rho}_{\rm eq}$ and ${\bar P}_{\rm eq}$ are the density and thermal pressure at equatorial plane. The scale-height of the accretion flow would then approximately be given by 
\begin{eqnarray}
h \sim \sqrt{\frac{c^2_s \, r^3}{ 1+  \frac{1}{8\pi \rho_0}  \left(B_{\varphi 0}  B_{\varphi 2} + B_{r0}  B_{r2} + \frac{9}{4} B_{r0}  B_{z1} \right) }  } \, .
\label{47} 
\end{eqnarray}
If we neglect the magnetic components, Eqn. (47) is then the  
usual hydrostatic scale-height of the accretion flow. Due to the presence of magnetic field, the scale-height is now approximately reduced by a factor 
$\sqrt{1+\frac{1}{8 \pi \rho_0} \left(B_{\varphi 0}  B_{\varphi 2} + B_{r0}  B_{r2} + \frac{9}{4} B_{r0}  B_{z1} \right)}$. 

As the region becomes more compressed, the thermal content of the gas increases and 
the excess thermal pressure gradient will help in lifting the plasma vertically outwards. 



\subsection{Case 2. For $\dot M = 10^{-3} \, \dot M_{\rm Edd} $ }

\begin{table*}
\large
\centerline{\large Table 2.1}
\centerline{\large $\dot M = 10^{-3} \dot M_{\rm Edd} $, $\beta = 0.9$, $f= 0.1$,  $\alpha = 0.3$, $t = 0.1$, $q=1.85$.}
\begin{center}
\begin{tabular}{cccccccc}
\hline
\hline
$\rho_0$  & $v_{r0}$  & $v_{\varphi 0}$ & $v_{z1}$  & $c_{s0}$  & $B_{r0}$ & $B_{\varphi 0}$ & $B_{z1}$ \\
\hline
8.0216e-4  &  -0.0900 & 0.5100  & 0.7370 & 0.5430 &  0.0028 & -0.0160 & 7e-4  \\
\hline
$\rho_2$  & $v_{r2}$  & $v_{\varphi 2}$ & $v_{z3}$  & $c_{s2}$  & $B_{r2}$ & $B_{\varphi 2}$ & $B_{z3}$ \\
\hline
-4.7e-3  &  -0.1960 & 0.2462 & -39.9113 &  -0.3761 & 0.1411 & -0.8040 & 0.1058 \\ 
\hline
\hline
\end{tabular}
\end{center}
\end{table*}

\begin{table*}
\large
\centerline{\large Table 2.2}
\centerline{\large $\dot M = 10^{-3} \dot M_{\rm Edd} $, $\beta = 0.9$, $f= 0.1$, $\alpha = 0.5$, $t = 0.1$, $q=1.85$.}
\begin{center}
\begin{tabular}{cccccccc}
\hline
\hline
$\rho_0$  & $v_{r0}$  & $v_{\varphi 0}$ & $v_{z1}$  & $c_{s0}$  & $B_{r0}$ & $B_{\varphi 0}$ & $B_{z1}$ \\
\hline
4.813e-4  &  -0.1498 & 0.5106  & 0.4168 & 0.5451 &  0.0040 & -0.0136 & 1e-3  \\
\hline
$\rho_2$  & $v_{r2}$  & $v_{\varphi 2}$ & $v_{z3}$  & $c_{s2}$  & $B_{r2}$ & $B_{\varphi 2}$ & $B_{z3}$ \\
\hline
-1.5e-3  &  -0.0939 & 0.2377 & -9.0662 &  -0.2749 & 0.0721 & -0.2418 & 0.0541 \\ 
\hline
\hline
\end{tabular}
\end{center}
\end{table*}

\begin{table*}
\large
\centerline{\large Table 2.3}
\centerline{\large $\dot M = 10^{-3} \dot M_{\rm Edd} $, $\beta = 0.9$, $f= 0.1$, $\alpha = 0.5$, $t = 0.2$, $q=1.85$.}
\begin{center}
\begin{tabular}{cccccccc}
\hline
\hline
$\rho_0$  & $v_{r0}$  & $v_{\varphi 0}$ & $v_{z1}$  & $c_{s0}$  & $B_{r0}$ & $B_{\varphi 0}$ & $B_{z1}$ \\
\hline
4.813e-4  &  -0.2948 & 0.5090  & 0.6330 & 0.5533 &  0.0127 & -0.0216 & 3.2e-3  \\
\hline
$\rho_2$  & $v_{r2}$  & $v_{\varphi 2}$ & $v_{z3}$  & $c_{s2}$  & $B_{r2}$ & $B_{\varphi 2}$ & $B_{z3}$ \\
\hline
-2.8e-3  &  -0.5535 & 0.5914 & -11.3549 &  -0.4235 & 0.1753 & -0.2907 & 0.1315 \\ 
\hline
\hline
\end{tabular}
\end{center}
\end{table*}

The accretion paradigm corresponding to this $\dot M$ resembles that in case 1. 
We choose similar values of $\beta$, $f$, $q$ and $t$ to study the 
feature of accretion-outflow coupled dynamics. We consider two values of 
$\beta$, $\beta \, = \, 0.95$ and $0.9$ corresponding to $q \, = \, 1.9$ and $1.85$ 
respectively, keeping the cooling factor $f$ same. This slightly less $q$ for 
$\beta \, = \, 0.9$ is ascribed to the fact that with the decrease in $\beta$, 
the content of the gas pressure in the system decreases, which makes the flow to 
be less sub-Keplerian. Other values of $\alpha$ and $t$ are same as in \S \RNum{4}(A)
We do not show the values of the coefficients for $\beta \, \sim \, 0.95$,  
$q \, = \, 1.9$ as they are very similar to that of the scenario for 
$\dot M \, = 10^{-4} \, \dot M_{\rm Edd}$, however, only present the 
values of the coefficients for $\beta \, \sim \, 0.9$,  $q \, = \, 1.85$ in tables 2.1, 2.2, 2.3. 
The `effective ratio of specific 
heat' $\Gamma$, corresponding to $\beta \, \sim \, 0.9$ is $\sim 1.61$. 
Resembling the scenario corresponding to $\dot M \, = 10^{-4} \, \dot M_{\rm Edd} $, here too, we get solutions 
only at high $\alpha$ and at a reduced $t$, the reason being argued in the previous subsection.

\subsection{Case 3. For $\dot M = 10^{-2} \, \dot M_{\rm Edd} $ }

\begin{table*}
\large
\centerline{\large Table 3.1}
\centerline{\large $\dot M = 10^{-2} \dot M_{\rm Edd} $, $\beta = 2/3$, $f= 0.5$,  $\alpha = 0.3$, $t = 0.05$, $q=1.75$.}
\begin{center}
\begin{tabular}{ccccccccc}
\hline
\hline
$\rho_0$  & $v_{r0}$  & $v_{\varphi 0}$ & $v_{z1}$  & $c_{s0}$  & $B_{r0}$ & $B_{\varphi 0}$ & $B_{z1}$ \\
\hline
0.0143  &  -0.0509 & 0.9211 & 0.6728 & 0.2462 &  0.0017 & -0.0293 & 4.25e-4  \\
\hline
$\rho_2$  & $v_{r2}$  & $v_{\varphi 2}$ & $v_{z3}$  & $c_{s2}$  & $B_{r2}$ & $B_{\varphi 2}$ & $B_{z3}$ \\
\hline
-0.3132  &  -0.4994 & -0.0714 & -152.2853 & -0.8796   & 0.3606 & -6.1736 & 0.2704 \\ 
\hline
\hline
\end{tabular}
\end{center}
\end{table*}

\begin{table*}
\large
\centerline{\large Table 3.2}
\centerline{\large $\dot M = 10^{-2} \dot M_{\rm Edd} $, $\beta = 2/3$, $f= 0.5$,  $\alpha = 0.5$, $t = 0.05$, $q=1.75$.}
\begin{center}
\begin{tabular}{ccccccccc}
\hline
\hline
$\rho_0$  & $v_{r0}$  & $v_{\varphi 0}$ & $v_{z1}$  & $c_{s0}$  & $B_{r0}$ & $B_{\varphi 0}$ & $B_{z1}$ \\
\hline
0.0086  &  -0.0847 & 0.9218  & 0.4130 & 0.2468 &  0.0024 & -0.0253 & 6e-4  \\
\hline
$\rho_2$  & $v_{r2}$  & $v_{\varphi 2}$ & $v_{z3}$  & $c_{s2}$  & $B_{r2}$ & $B_{\varphi 2}$ & $B_{z3}$ \\
\hline
-0.1119  &  -0.4971 & -0.1303 & -38.7978 & -0.5916 & 0.1930 & -2.0099 & 0.1447 \\ 
\hline
\hline
\end{tabular}
\end{center}
\end{table*}

\begin{table*}
\large
\centerline{\large Table 3.3}
\centerline{\large $\dot M = 10^{-2} \dot M_{\rm Edd} $, $\beta = 2/3$, $f= 0.5$,  $\alpha = 0.5$, $t = 0.1$, $q=1.75$.}
\begin{center}
\begin{tabular}{ccccccccc}
\hline
\hline
$\rho_0$  & $v_{r0}$  & $v_{\varphi 0}$ & $v_{z1}$  & $c_{s0}$  & $B_{r0}$ & $B_{\varphi 0}$ & $B_{z1}$ \\
\hline
0.0086  &  -0.1689 & 0.9244  & 0.6179 & 0.2492 &  0.0084 & -0.0441 & 2.1e-3  \\
\hline
$\rho_2$  & $v_{r2}$  & $v_{\varphi 2}$ & $v_{z3}$  & $c_{s2}$  & $B_{r2}$ & $B_{\varphi 2}$ & $B_{z3}$ \\
\hline
-0.1931  &  -1.1612 & -0.3876 & -46.8950 & -0.9662 & 0.5086 & -2.6172 & 0.3815 \\ 
\hline
\hline
\end{tabular}
\end{center}
\end{table*}

\begin{table}
\Large
\centerline{\large Table 4}
\centerline{\large $\beta = 2/3$, $f= 0.4$}
\begin{center}
\begin{tabular}{ccccccc}
\hline
\hline
$\alpha$  & $t$  & $v_{\varphi 2}$ \\
\hline
0.3  & 0.05 & -0.0202  \\
\hline
0.5  & 0.05 & -0.0485 \\
\hline
0.5  & 0.1 & -0.1818 \\
\hline
\hline
\end{tabular}
\end{center}
\end{table}

The earlier values of $\dot M$ in the previous subsections correspond 
to RIAF. Nonetheless, moderately advective accretion paradigm may also be 
susceptible to eject outflows and jets. These accretion flows which are less advective 
as compared to RIAFs, will have lesser gas pressure 
content and higher cooling efficiency, and are less geometrically thick and are centrifugally more dominating. 
They have a moderate optical depth. This accretion paradigm can be presumably envisaged with 
$10^{-3} \dot M_{\rm Edd} << \dot M \, \lsim \, 10^{-2} \, \dot M_{\rm Edd}$; for our analysis, here, 
we choose $\dot M \, \sim 10^{-2} \, \dot M_{\rm Edd} $. 

We choose appropriate values of $\beta$ and $f$ corresponding to this $\dot M$ to  
get physical plausible solutions. As this flow is sub-Keplerian
and advective, $q$ should be greater that $1.5$, but considerably less than 
in flows illustrated in previous subsections. We choose the value 
of $q \sim 1.75$. For $1.5 < q \, << \, 1.75$ and for $1.75 \, << \, q < 1.85$, 
we never found any physically valid solutions with $\dot M \, \sim 10^{-2} \, \dot M_{\rm Edd}$. 
However, in the stated range one may still obtain valid solutions, either for $\dot M \, >> \, 10^{-2} \, \dot M_{\rm Edd}$ or 
for $\dot M \, << \, 10^{-2} \, \dot M_{\rm Edd}$, respectively.  

We evaluate the value of the dynamical 
coefficients for appropriate choice of $\beta \sim  2/3$ with 
$f = 0.4, \, 0.5$ corresponding to $\dot M \sim 10^{-2} \, \dot M_{\rm Edd} $ . The values of $\alpha$ are same as before. We found 
that for $f\, > \, 0.5$ or $\beta\, \lsim \, 0.6$, the flow becomes near Keplerian 
($v_{\varphi 0} \sim 1$) and we get the physical solutions only at 
$t \,< \,0.05$. Such a flow is not favourable for ejection of 
outflow as reasoned earlier. Hence we restrict our study to 
a maximum value of $f = 0.5$ corresponding to 
$\beta \sim  2/3$. Corresponding $\Gamma$ for  $\beta \sim  2/3$ is $1.5$. 
In tables 3.1, 3.2, 3.3, we present them for $f = 0.5$ for  
appropriate values of $t$. 

For an easy comparison we furnish the values of 
$v_{\varphi 2}$ corresponding to $f=0.4$ for $\dot M \sim 10^{-2} \, \dot M_{\rm Edd}$, 
with appropriate $\alpha$ and $t$ in Table 4, whose importance 
we will notice as we proceed. 

We found that we do not obtain any solution for $t\, > \,0.05$ with  
$\alpha = 0.3$, and $t\, > \,0.1$ for $\alpha = 0.5$. The reason of physical solutions 
of accretion-induced outflow at a reduced scale-height has been stated in \S \RNum{4}(A). However, obtaining 
physically valid solutions 
with $\dot M \sim 10^{-2} \, \dot M_{\rm Edd}$ at a slightly reduced scale-height as compared to that 
obtained with $\dot M \, \lsim \, 10^{-3} \, \dot M_{\rm Edd}$ is consistent with the fact, that 
as $ \dot M > 10^{-3} \, \dot M_{\rm Edd}$, the accretion flow tends to become more rotationally dominated 
with diminishing degree of advection.   

If we compare the value of the dynamical coefficients for 
$ \dot M \, \lsim \, 10^{-3} \, \dot M_{\rm Edd}$ and $ \dot M \, \sim \, 10^{-2} \, \dot M_{\rm Edd} $, 
we found a fundamental difference in the dynamical nature of $\bar v_{\varphi}$. For
$\dot M \, \lsim \, 10^{-3} \, \dot M_{\rm Edd}$, the value of the coefficient 
$v_{\varphi 2}$ is always positive. On the contrary, the value of $v_{\varphi 2}$ is negative 
for  $\dot M \sim 10^{-2} \, \dot M_{\rm Edd}$. Negative value of $v_{\varphi 2}$ implies 
that $\bar v_{\varphi}$ decreases in $z$ in the coupled accretion outflow region for   
$\dot M \sim 10^{-2} \, \dot M_{\rm Edd}$. To verify this anomaly, we evaluated 
$v_{\varphi 2}$ for lower $\beta$ and 
higher $f$, and vice-versa. It is revealing 
that for $f > 0.3$ and $\beta < 0.75$, we always obtain a negative 
$v_{\varphi 2}$ with $\alpha \sim 0.5$. $v_{\varphi 2} = 0.0162$ corresponding to 
$f = 0.3$, $\beta = 0.75$. With $\alpha \sim 0.3$ the negative values of 
$v_{\varphi 2}$ are obtained for $f \, \gsim \, 0.4$ and $\beta < 0.7$. The 
corresponding value of $v_{\varphi 2} = 0.0275$ for 
$f = 0.4$ and $\beta = 0.7$. To reassure ourselves we computed 
$v_{\varphi 2}$ for $\dot M \sim 10^{-1} \, \dot M_{\rm Edd}$, and we arrive at a similar 
result. The above consistent 
findings convey that for strong gas pressure and advection dominated 
flows (RIAFs), $\bar v_{\varphi}$ do 
not decrease in $z$ within the accretion-outflow coupled region. On the 
contrary, for flows with lesser content of gas and higher cooling efficiency, 
which are less advective and centrifugally more dominating, $\bar v_{\varphi}$ decreases 
in $z$ within the coupled accretion-induced outflow region. We comment on this 
apparent dichotomy in \S \RNum{5}. We also found that with a moderate 
decrease in $\alpha$ from $0.5$ to $0.3$, $v_{\varphi 2}$ turns negative at a higher 
$f$ and at a lower value of $\beta$, corresponding to  $\dot M \sim  10^{-2} \, \dot M_{\rm Edd}$. 

\section{Dynamics and nature of the magnetized accretion-induced outflow}

In this section, we analyze the family of solutions for advective flows 
in the accretion-outflow coupled region with 
$\dot M \, \lsim \, 10^{-2} \, \dot M_{\rm Edd}$. Although the flow 
variables vary in both $r$ and $z$, we do not display  
three dimensional figures as they are very obscure and difficult to 
interpret. As outflow and jet effuse out 
from inner region of the accretion flow, we restrict our analysis up to 
$50$ Schwarzschild radius ($r_g$) within which we presume that the accretion 
and outflow are coupled, where, $r_g = 2GM/c^2$. Also it has been stated by Kumar et al. (2013) that 
VLBI observations of M87 (Junor et al. 1999) have shown that the jet originates from the vicinity ($\sim 50 r_g$) of a 
BH/compact object. However, we 
do not expect any outflow out of the accreting plasma in the 
extreme vicinity of a BH, as the accretion flow 
remains highly bounded in the extreme vicinity of a 
gravitationally starved BH. So we restrict our study up to $5 \, r_g$ in the inward radial direction, a quite 
reasonable choice for the inner radius of the accretion-outflow coupled region. 
Both $r$ and $z$ coordinates in the figures are expressed in 
units of $r_g$. We express $\dot M$ in our entire 
analysis in units 
of $\dot M_{\rm Edd}$. The dynamical solutions are shown in the 
following figures 
with $G = M = 1$. Note that all the dynamical variables in our 
study are mean quantities, which are either time averaged or 
ensemble averaged. All the flow variables in the figures represent 
mean quantities. 

Figures 1 and 2 describe the variation of 
vertically averaged flow variables as functions of radial coordinate 
$r$ in different accretion paradigms. Variation 
of $\bar v_z$ and $\bar B_z$ are shown only along coupled 
accretion-outflow surface $h(r)$ as they are odd functions in $z$. 
Figure 1 shows that with the decrease in $\dot M$, the poloidal 
components of the velocity ($\bar v_r, \bar v_z$) and the sound speed 
consistently increases. Conversely, 
the magnitude of $\bar v_{\varphi}$ increases with the flow becoming less 
advective and more centrifugally dominated. It is being interestingly found from the tables in \S \RNum{4}, 
that the value of the large-scale poloidal magnetic field enhances with the 
increase in the geometrical thickness of the accretion flow. As one moves from strongly advective regime to 
moderately advective regime, there is a sharp fall in the value of the poloidal component of the magnetic field. 
This is owing to the fact that the geometrical thickness of the accretion region corresponding to 
moderately advective accretion flow is much less as compared to the case in strongly advective accretion paradigm, as shown 
in figures 2a,b. This indicates the dominating 
influence of the vertical thickness of the accretion flow structure 
on the poloidal component of the magnetic field. In contrast, the toroidal component of the 
magnetic field $\bar B_{\varphi}$ always increases with the flow becoming less advective 
and more centrifugally/rotationally dominated (Fig. 2c). 

Figure 3 shows the variation of 
poloidal component of velocity and magnetic field with $\alpha$ 
for different accretion paradigms along the radial distance $r$. $\bar v_r$ and 
$\bar B_r$ are vertically averaged quantity, whereas 
$\bar v_z$ and $\bar B_z$ are along coupled accretion-outflow surface $h(r)$. 
We choose the value of corresponding $t$ for different 
$\alpha$ to be maximum as illustrated in \S \RNum{4}. This is because 
the system has a greater tendency 
to relax itself to the maximum possible height available to render 
a physical plausible solution of the coupled accretion-outflow, as 
the geometrically thicker accretion flow is more conducive to propel 
plasma vertically outwards of the accretion region. We find that with 
a small increase in $\alpha$ from $0.3$ to $0.5$, the value of the poloidal component 
of velocity and the magnetic field increases for both accretion paradigms. We do not 
show the variation of other flow variables with $\alpha$ as their 
dependence on $\alpha$ is insignificant for a particular $\dot M$, which can be 
verified from tables in \S \RNum{4}.  

\begin{figure*}
\centering
\includegraphics[width=170mm]{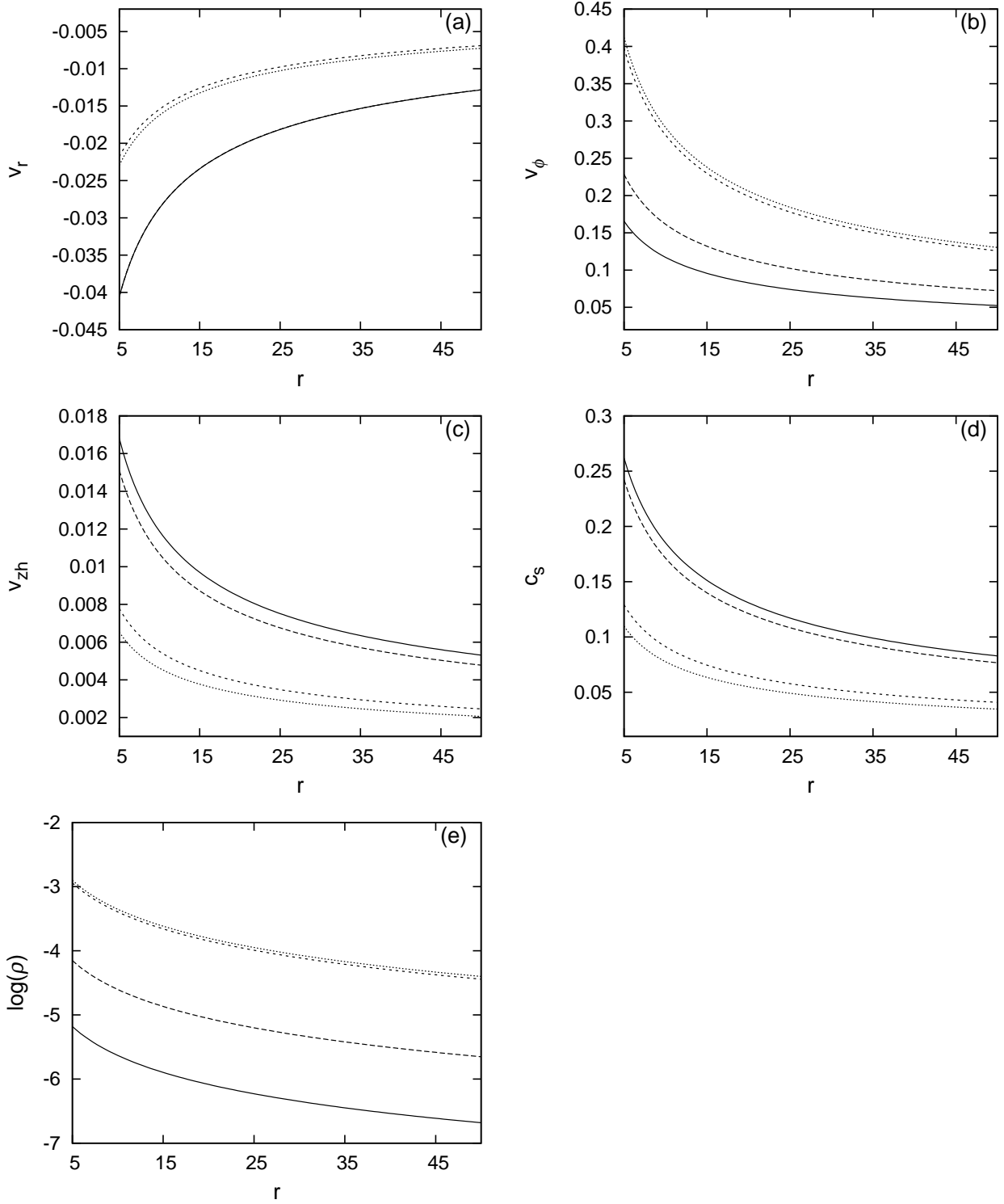}
\caption{Variation of (a) vertically averaged radial velocity, (b) vertically averaged toroidal velocity, (c) vertical/outflow 
velocity at height $h$, (d) vertically averaged sound speed, (e) vertically averaged density 
along radial coordinate $r$. $r$ is expressed 
in units of Schwarzschild radius. Solid, long-dashed, short-dashed 
and dotted curves are for ($\dot M = 10^{-4}, 10^{-3}, 10^{-2}, 10^{-2}$) 
with corresponding ($f$, $\beta$) $=$ ($0.1,0.95$), ($0.1,0.9$), ($0.4,2/3$), 
($0.5,2/3$), respectively. Other parameter is $\alpha=0.3$. The flow variables in the vertical axis are 
in units of $\sqrt{GM/{r_g}}$, and density in units of 
${(GM)^{-1/2} \times \dot M_{\rm Edd}}/{r^{3/2}_g} $. $\dot M$ is expressed in units of Eddington accretion rate.  
 }
\label{Fig1}
\end{figure*}

\begin{figure*}
\centering
\includegraphics[width=170mm]{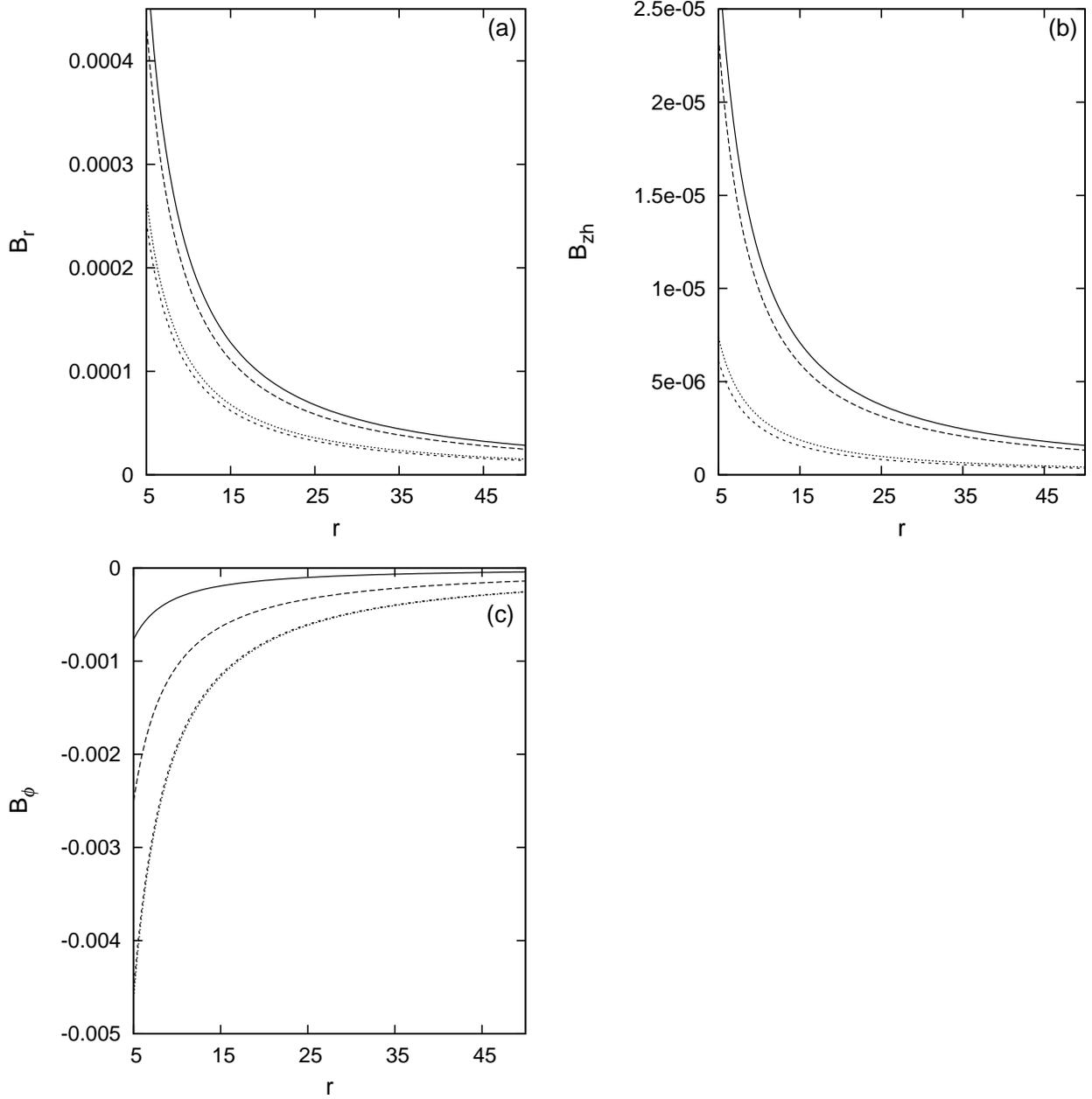}
\caption{Variation of (a) vertically averaged radial magnetic field, 
(b) vertically averaged toroidal magnetic field, (c) vertical magnetic field at
$h$, along radial coordinate $r$. Solid, long-dashed, short-dashed 
and dotted curves are for ($\dot M = 10^{-4}, 10^{-3}, 10^{-2}, 10^{-2}$) 
with corresponding ($f$, $\beta$) $=$ ($0.1,0.95$), ($0.1,0.9$), ($0.4,2/3$), 
($0.5,2/3$), respectively. Other parameter is $\alpha=0.3$. $r$ is expressed in units of Schwarzschild radius. Magnetic fields are in units of 
${(\sqrt{GM} \times \dot M_{\rm Edd})^{1/2}}/{r^{5/4}_g} $. $\dot M$ is expressed in units of Eddington accretion rate.
 }
\label{Fig2}
\end{figure*}

\begin{figure*}
\centering
\includegraphics[width=170mm]{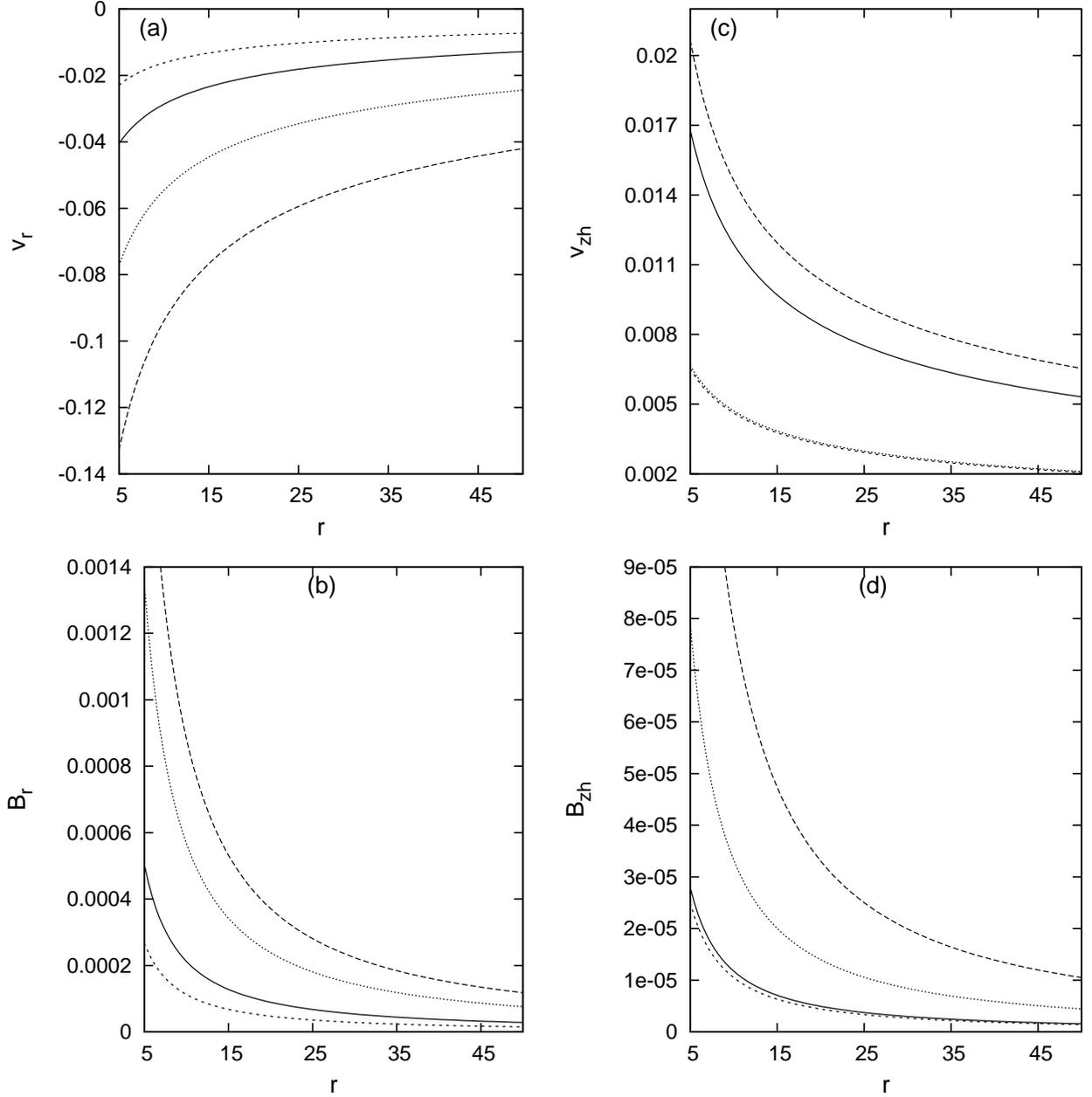}
\caption{Variation of poloidal components of velocity and magnetic field with $\alpha$ 
for two different accretion paradigms along $r$. Solid, long-dashed, short-dashed and dotted curves 
are for ($\dot M, \alpha, f, t$) = ($10^{-4}, 0.3, 0.1, 0.1$), 
($10^{-4}, 0.5, 0.1, 0.2$), ($10^{-2}, 0.3, 0.5, 0.05$), ($10^{-2}, 0.5, 0.5, 0.1$).
The units of the variables along the axes are same as in figures 1 and 2. $\dot M$ is expressed in units of Eddington accretion rate.
 }
\label{Fig3}
\end{figure*}

\begin{figure*}
\centering
\includegraphics[width=170mm]{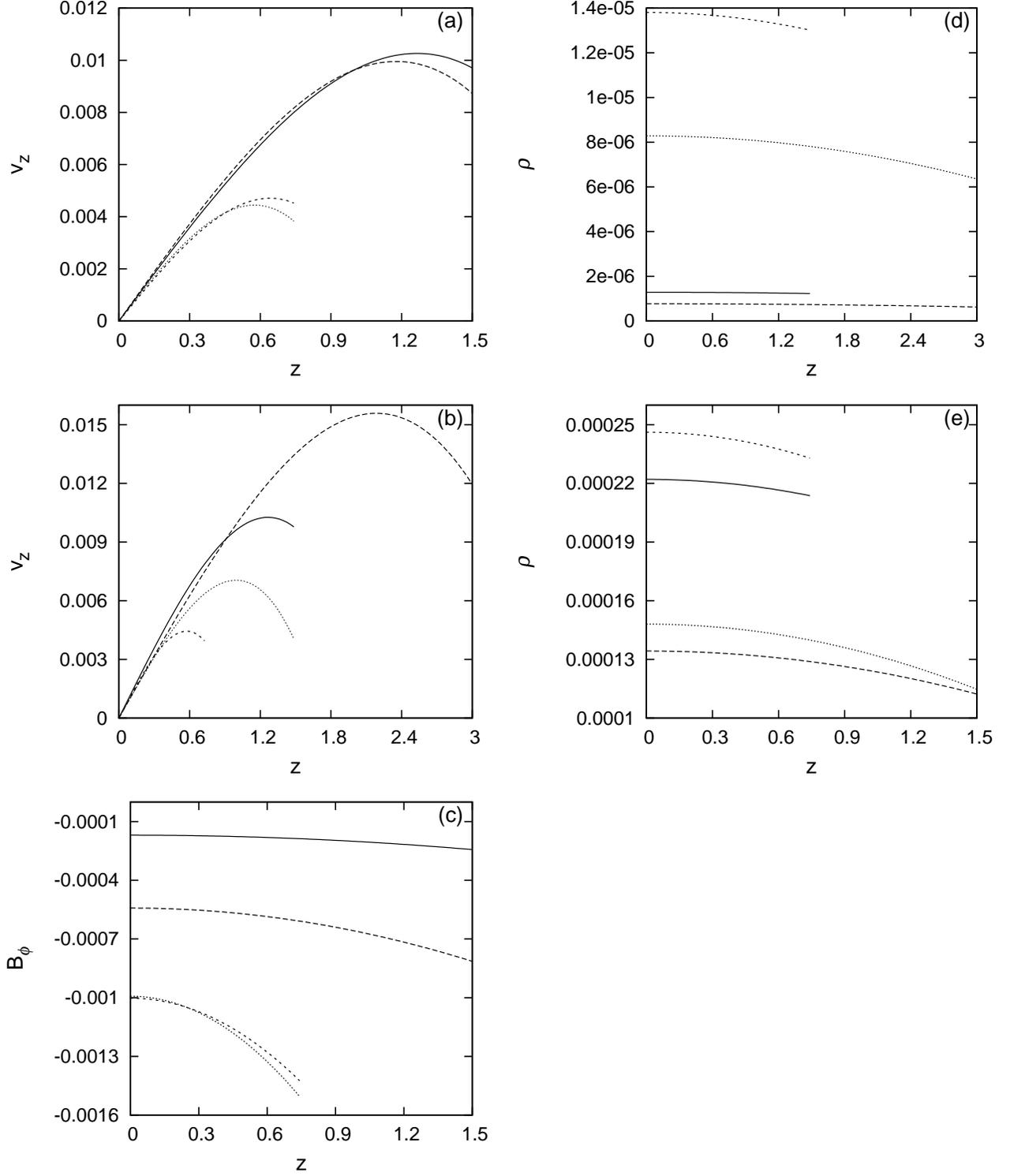}
\caption{Variation of flow variables along vertical coordinate $z$ corresponding 
to $r=15 \, r_g$. 
Solid, long-dashed, short-dashed and dotted curves in (a), (b) and 
(c) are for same parameters corresponding to figures 1, 3 and 1 
respectively. Solid, long-dashed, short-dashed and dotted curves 
in (d) are for ($\dot M, \alpha, \beta, f$) = ($10^{-4}, 0.3, 0.95, 0.1$), 
($10^{-4}, 0.5, 0.95, 0.1$), ($10^{-3}, 0.3, 0.9, 0.1$), 
($10^{-3}, 0.5, 0.9, 0.1$). Similarly the corresponding curves in (e) are 
for ($10^{-2}, 0.3, 2/3, 0.4$), ($10^{-2}, 0.5, 2/3, 0.4$), 
($10^{-2}, 0.3, 2/3, 0.5$), ($10^{-2}, 0.5, 2/3, 0.5$). 
The units of the variables along the axes are same as in figures 1 and 2
and 3. $\dot M$ is expressed in units of Eddington accretion rate.
 }
\label{Fig4}
\end{figure*}

\begin{figure*}
\centering
\includegraphics[width=170mm]{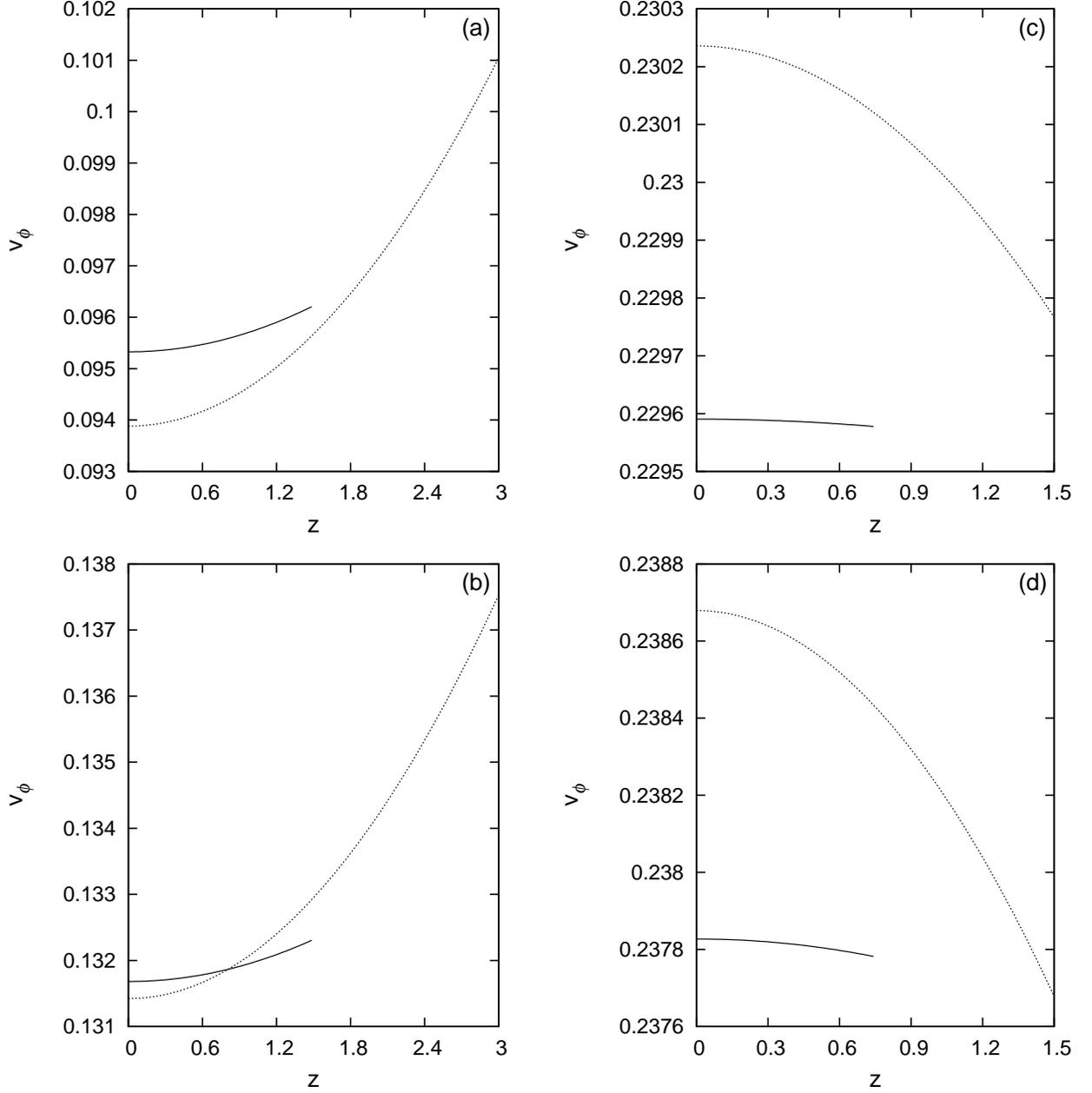}
\caption{Variation of toroidal velocity as a function of vertical 
coordinate $z$ corresponding to $r=15 \, r_g$. The 
solid and dotted curves are for ($\dot M, \alpha, \beta,f$) in 
(a) = ($10^{-4}, 0.3, 0.95, 0.1$), ($10^{-4}, 0.5, 0.95, 0.1$), 
(b) = ($10^{-3}, 0.3, 0.9, 0.1$), ($10^{-3}, 0.5, 0.9, 0.1$),  
(c) = ($10^{-2}, 0.3, 2/3, 0.4$), ($10^{-2}, 0.5, 2/3, 0.4$),
(d) = ($10^{-2}, 0.3, 2/3, 0.5$), ($10^{-2}, 0.5, 2/3, 0.5$).
The units of variables along the axes are same as in earlier figures. 
$\dot M$ is expressed in units of Eddington accretion rate.
 }
\label{Fig5}
\end{figure*}

\begin{figure*}
\centering
\includegraphics[width=170mm]{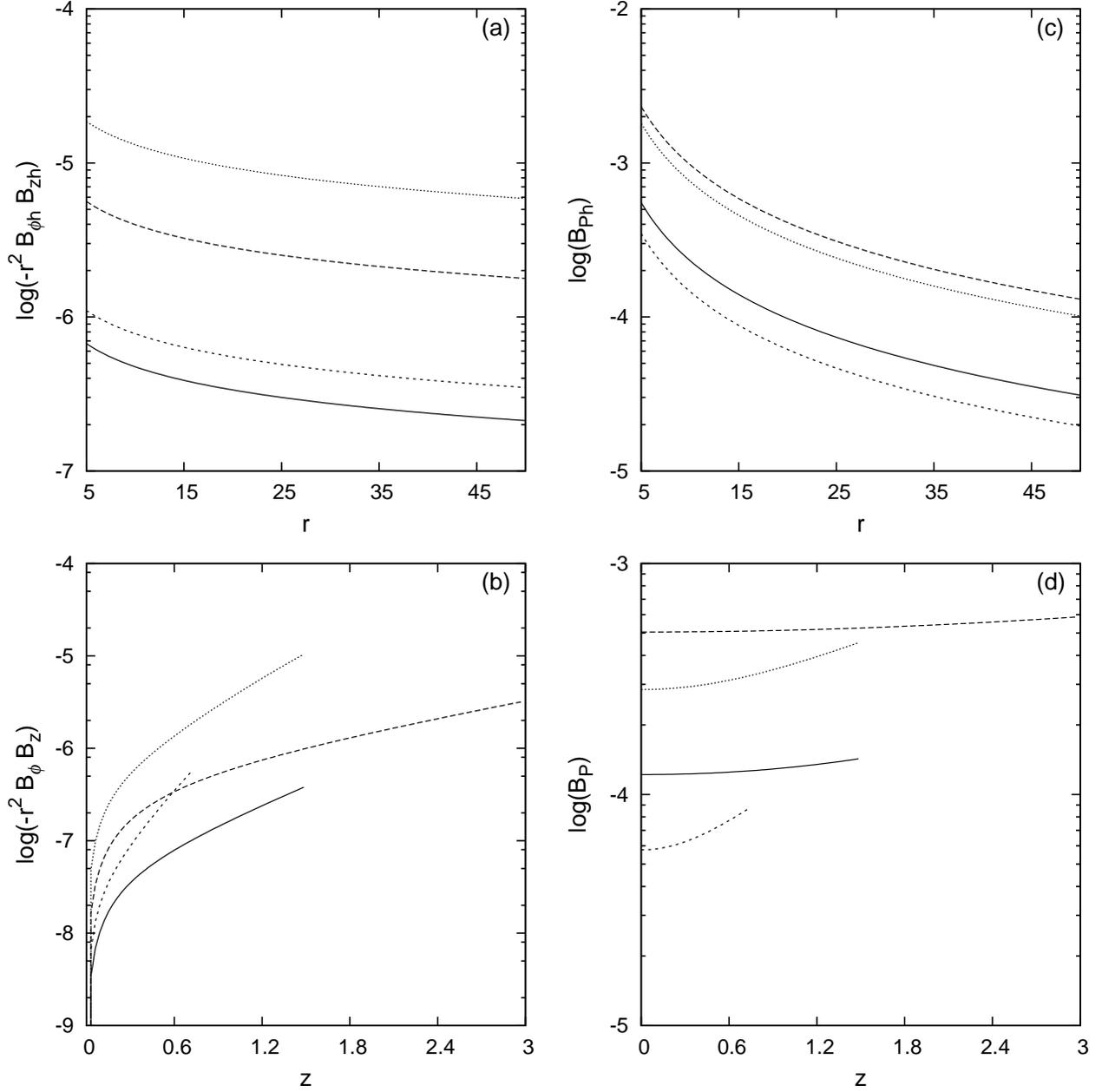}
\caption{Variation of differential magnetic torque and poloidal magnetic field 
along $r$ and $z$. Variation along $z$ is at $r=15 \, r_g$.  
Solid, long-dashed, short-dashed and dotted curves in (a), (b), (c) and
(d) are for ($\dot M, \alpha, f, t$) = ($10^{-4}, 0.3, 0.1, 0.1$), 
($10^{-4}, 0.5, 0.1, 0.2$), ($10^{-2}, 0.3, 0.5, 0.05$), ($10^{-2}, 0.5, 0.5, 0.1$). 
The component of magnetic fields are expressed in units similar to earlier figures. $r$ and $z$ 
are expressed in units of Schwarzschild radius. 
$\dot M$ is expressed in units of Eddington accretion rate.
 }
\label{Fig6}
\end{figure*}

\begin{figure*}
\centering
\includegraphics[width=170mm]{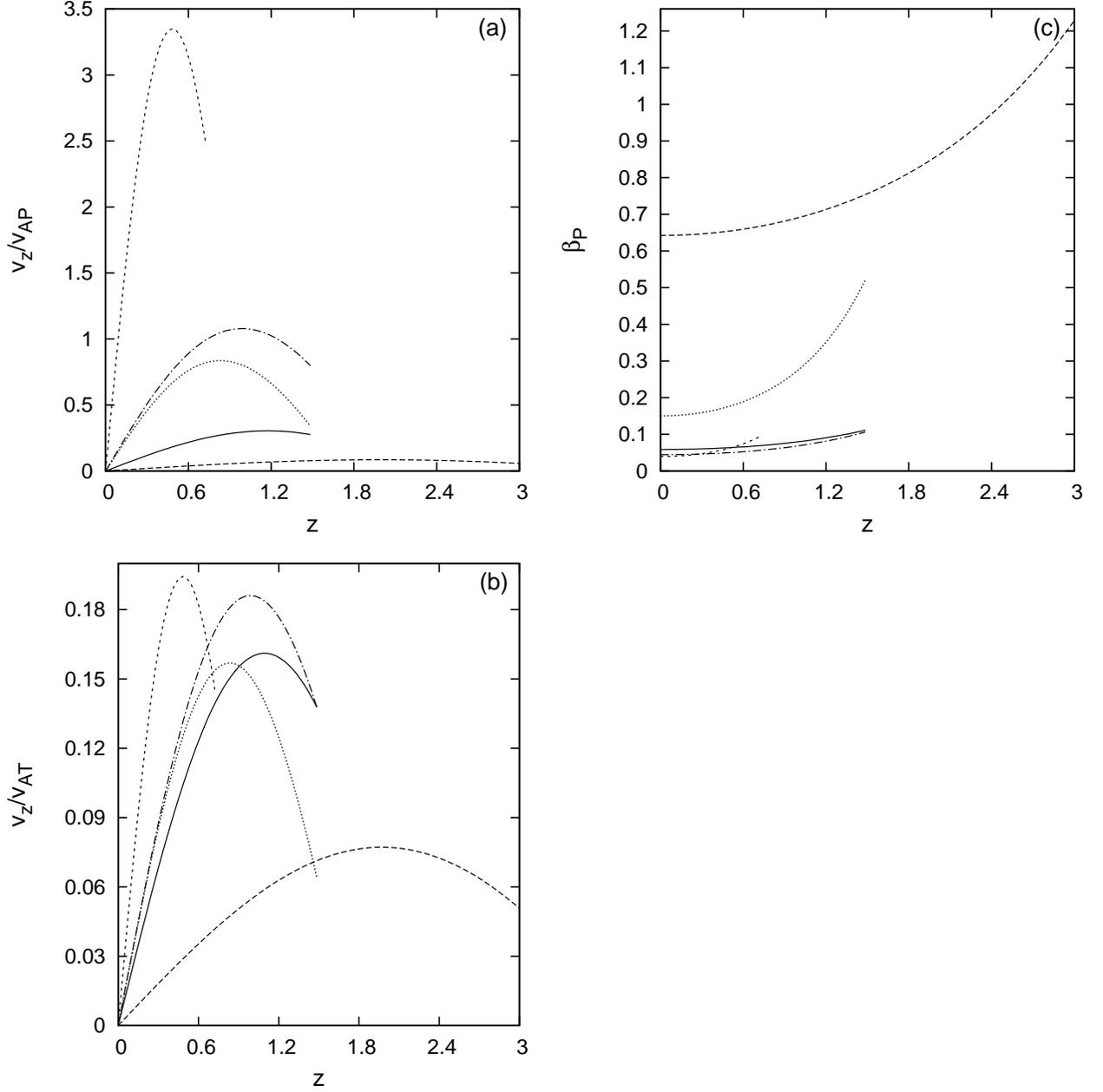}
\caption{Variation of the (a) ratio of vertical velocity to poloidal Alfven 
velocity, (b) ratio of vertical velocity to total Alfven velocity, 
(c) plasma $\beta_P$ along $z$ corresponding to $r=15 \, r_g$. 
Solid, long-dashed, short-dashed, dotted, dot-dashed curves in 
(a), (b) and (c) are for ($\dot M, \alpha, f, t$) = 
($10^{-4}, 0.3, 0.1, 0.1$), 
($10^{-4}, 0.5, 0.1, 0.2$), 
($10^{-2}, 0.3, 0.5, 0.05$), 
($10^{-2}, 0.5, 0.5, 0.1$)
($10^{-3}, 0.3, 0.1, 0.1$), 
 $\dot M$ is expressed in units of Eddington accretion rate.
 }
\label{Fig7}
\end{figure*}

\begin{figure*}
\centering
\includegraphics[width=170mm]{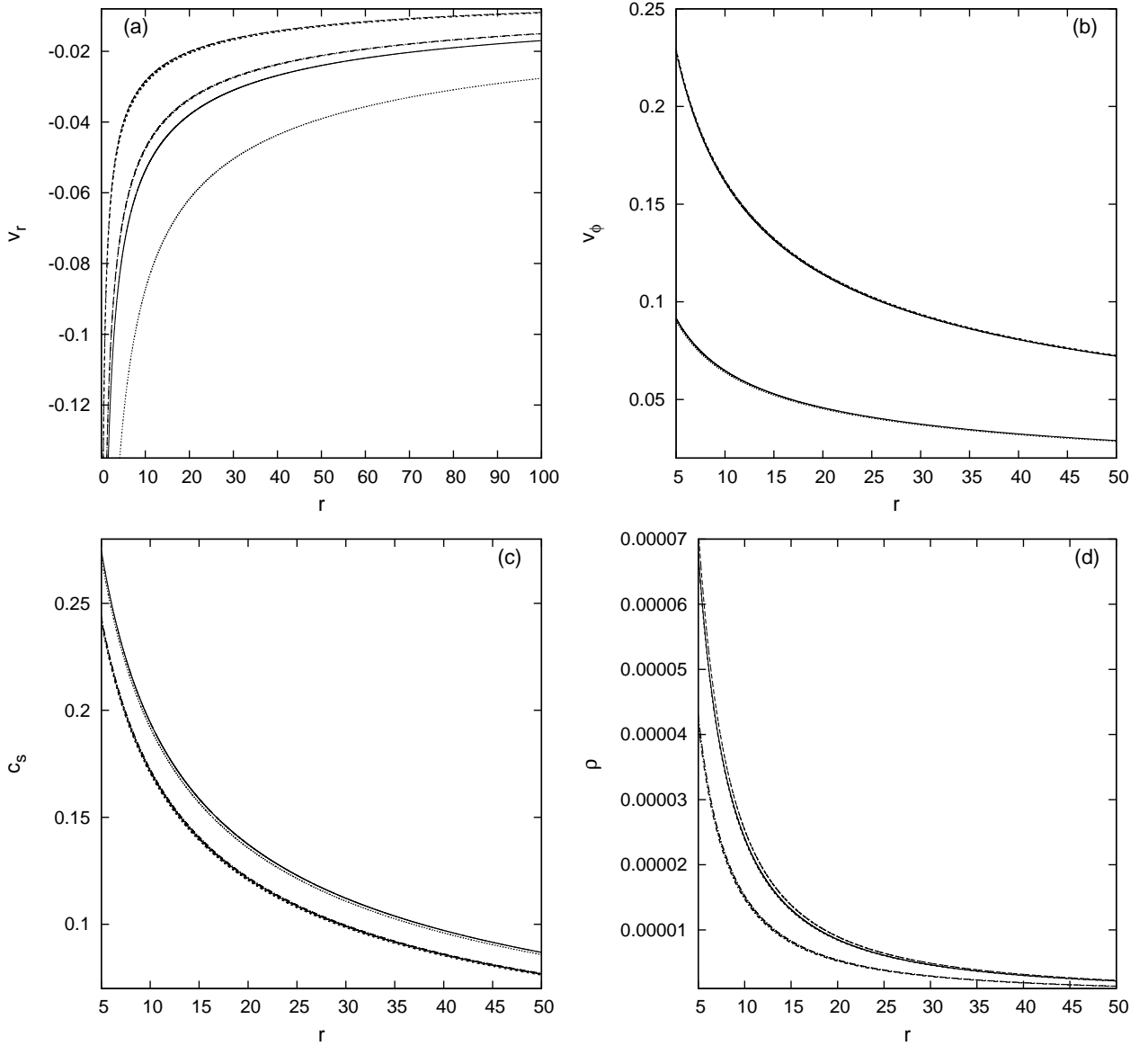}
\caption{Comparison of our magnetized accretion-outflow solutions with that of ADAF type solutions in Narayan \& Yi 1994 [5]. In 
figures 8a,b,c,d we compare radial velocity, orbital velocity, sound speed and density, respectively, obtained for our 
magnetized accretion-outflow solutions with that of ADAF type solutions. The curves correspond to $\dot M = 10^{-3}$. 
Solid, long-dashed, short-dashed curves in all the figures corresponding to $\alpha = 0.3$. 
In all the figures, solid and long-dashed curves are for flow variables at equatorial plane corresponding to ADAF type and for our 
magnetized accretion-outflow solutions, respectively. Short-dashed curves correspond to flow variables 
for our magnetized accretion-outflow solution at height $h$. Long-dashed curves and short-dashed curves correspond to $t= 0.1$. 
Dotted, long dotted-dashed, short dotted-dashed curves exactly resemble solid, long-dashed and short-dashed curves, however
correspond to $\alpha = 0.5$. Long dotted-dashed and short dotted-dashed curves correspond to equatorial plane and at scale-height 
$h$ for our magnetized accretion-outflow solution, corresponding to $t= 0.1$. 
Other parameters are ($\beta = 0.9, f= 0.1$). $\dot M$ is expressed in units of Eddington accretion rate. Velocities 
and density are expressed in units already stated in the caption in Fig. 1. 
 }
\label{Fig8}
\end{figure*}

\begin{figure*}
\centering
\includegraphics[width=170mm]{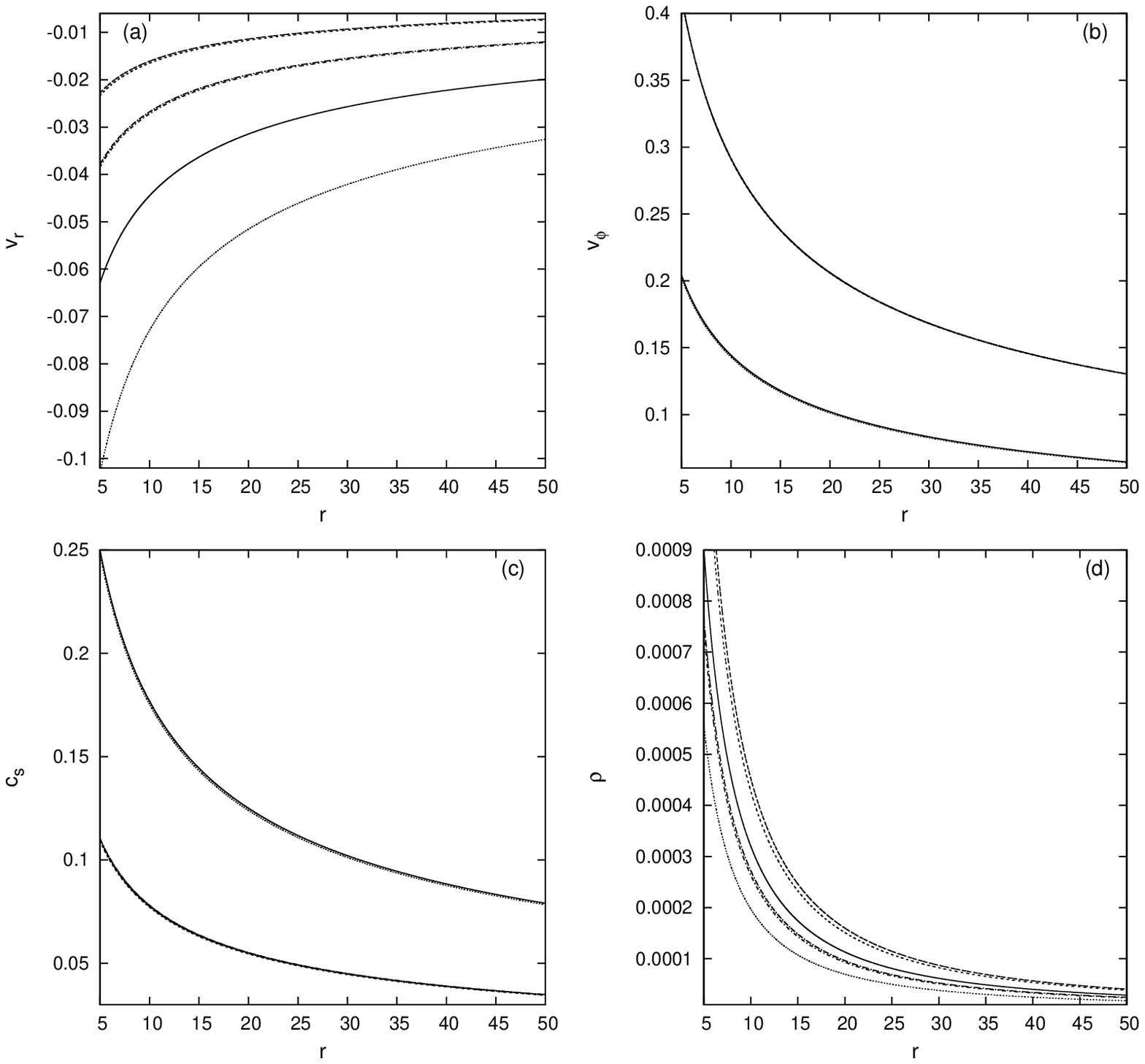}
\caption{Exactly similar to that of Fig. 8, however for  $\dot M = 10^{-2}$. 
Solid, long-dashed, short-dashed curves in all the figures corresponding to $\alpha = 0.3$. Solid and long-dashed 
curves are for flow variables at equatorial plane corresponding to ADAF type and for our 
magnetized accretion-outflow solutions, respectively. Short-dashed curves correspond to flow variables 
for our magnetized accretion-outflow solution at height $h$. Long-dashed and short-dashed curves correspond to 
$t= 0.05$. Dotted, long dotted-dashed, short dotted-dashed curves resemble solid, long-dashed and short-dashed curves, however
correspond to $\alpha = 0.5$. Long dotted-dashed and short dotted-dashed curves correspond to $t= 0.05$. 
Other parameters are ($\beta = 2/3, f= 0.5$). $\dot M$ is expressed in units of Eddington accretion rate. Velocities 
and density are expressed in units already stated in the caption in Fig. 1. 
 }
\label{Fig9}
\end{figure*}     

In figures 4 and 5 we present the variation of the few dynamical variables in 
$z$ (which are very relevant to outflow) at any arbitrary location in $r$, for 
$10^{-4} \, \dot M_{\rm Edd} \, \lsim \, \dot M \, \lsim \, 10^{-2} \, \dot M_{\rm Edd}$. 
Figure 4a shows the dependence 
of $\bar v_z$ in $z$ for different $\dot M$ for 
$\alpha = 0.3$ similar to that in Fig. 1. 
It is found that initially $\bar v_z$ increases rapidly in 
$z$, however there is a sudden deceleration of $\bar v_z$ as the flow 
approaches the coupled accretion-induced outflow surface. 
This is due to the fact that the inward vertical component of the gravitational force $(F_{Gz})$ dominates 
near the coupled accretion-outflow surface. Also at low $\dot M$ corresponding to 
moderately advective accretion flow, the increase 
of $\bar v_z$ in the vertical direction is much steeper. The truncation of the 
curves at a particular $z$ represent the corresponding vertical 
thickness $t$. Figure 4b shows the variation of $\bar v_z$ in $z$ with $\alpha$ for 
two different accretion paradigms similar to that in Fig. 3. The nature of the curves 
are similar to that in Fig. 4a. Nevertheless, with the increase 
in $\alpha$, the value of $\bar v_z$ gets enhanced, 
and there is a steeper increase of $\bar v_z$ in $z$. The nature 
of the variation of toroidal component of the magnetic field in $z$ for different $\dot M$ 
is shown in Fig. 4c. It is seen that with the increase in $\dot M$, 
$\bar B_{\varphi}$ increases at a much faster rate in $z$. Variation of 
$\bar B_{\varphi}$ with $\alpha$ is 
insignificant (see tables in \S \RNum{4}) and hence is not graphically displayed. 
In figures 4d and 4e, we show the variation of density in $z$ corresponding to strongly 
advective and moderately advective accretion regimes, respectively. Both the  
figures indicate that $\bar \rho$ decreases with the increase in 
$\alpha$ for all $\dot M$. Also, there is a steeper fall of density in $z$ 
with the increase in $\dot M$. Figure 5 depicts the variation of toroidal velocity $\bar v_{\varphi}$ in 
$z$. With the increase in $\dot M$ ($\dot M > 10^{-3} \, \dot M_{\rm Edd}$)   as the 
flow becomes less advective and more centrifugally dominated, $\bar v_{\varphi}$ 
decreases in $z$ for all relevant values of $\alpha$ and $f$. This feature 
of $\bar v_{\varphi}$ has already been remarked in \S \RNum{4}. In figures 5c and 5d, we show them 
for $\dot M \sim 10^{-2} \, \dot M_{\rm Edd}$. We also find that 
with increase in $\alpha$ the profile of $\bar v_{\varphi}$ in $z$ attains 
a steeper nature. 

In figures 6b and 6a we depict the profile of differential 
magnetic torque ($-r^2 \bar B_{\varphi} \bar B_z$) along $z$, as well as differential magnetic torque acting on the 
coupled accretion-outflow 
surface $\left(-r^2 \bar B_{\varphi h} \bar B_{zh} \right)$ along $r$, respectively, corresponding to different accretion 
paradigms. This term represents the 
magnetic extraction of angular momentum by the outflowing matter, what is called 
the magnetic braking. The curves show that with the increase in the value of $\dot M$ 
as well as with the increase in $\alpha$ (for a specific value of $\dot M$), the value of the differential 
magnetic torque increases. As $\dot M$ increases, 
with the accretion flow becoming less advective 
and more centrifugally dominated, the extraction 
of angular momentum by the outflowing plasma is greatly enhanced. Consequently the gas gets centrifugally accelerated and would 
leave the accretion region by removing the angular momentum from the accreting matter. This mechanism predominantly determines 
the outward flow of the matter for accretion flow with moderate advection. If 
the system has a large residual toroidal velocity (centrifugally dominated), 
it is possible that the angular momentum loss in the vertical 
direction due to the magnetic torque 
will be so high, owing to which the degree of angular momentum 
loss proportionately increases as the matter flow vertically outwards. As 
a consequence, there will be an eventual decrease of the toroidal velocity 
$\bar v_{\varphi}$ in the vertical direction within the accretion-outflow coupled 
region, as seen in Fig. 5c,d, corresponding to accretion flow with 
$\dot M \sim 10^{-2} \dot M_{\rm Edd}$. On the other hand, if the accretion flow is 
predominantly gas pressure dominated as in a strongly advective regime 
(with $\dot M \, \lsim \, 10^{-3} \, \dot M_{\rm Edd}$), the gas pressure gradient would play a more 
contributory role to lift the plasma vertically outwards with the 
help of magnetic forces, and the effective contribution of the  magnetocentrifugal acceleration 
to control the dynamics of the outflowing matter gets curtailed as compared to that in a more 
centrifugally dominated accreting system.

Figures 6d and 6c show the variation of 
the poloidal component of the magnetic 
field $\bar B_P$ [$ = \sqrt({\bar B}^2_r + {\bar B}^2_z)$] along $z$ and at coupled accretion-outflow surface 
$\left(\bar B_{Ph}\right)$ along $r$, 
respectively. The nature of the curves indicate that as high $\alpha$ 
renders the accretion-induced outflow to a greater geometrical thickness 
(commented earlier), the large-scale poloidal field 
gets strongly augmented with the increase in $\alpha$ for a specific 
$\dot M$ due to the dominating influence of the vertical thickness on $\bar B_P$. Thus 
with the increase in the turbulent viscosity parameter 
$\alpha$, the value of the differential magnetic torque responsible 
for the centrifugal acceleration of the outflowing plasma, as well as the 
large-scale poloidal magnetic field $\bar B_P$ get strongly augmented; consequently enhancing 
the transport of vertical flux outwards. In the moderately advective accretion paradigm with more centrifugal 
domination, the effective contribution to launch and eject the matter vertically 
outwards from the accretion region arises mainly from 
the magnetocentrifugal acceleration. A small increase in the turbulent viscosity
parameter $\alpha$ from $0.3$ to $0.5$ intensifies the process
of extraction of the angular momentum due to the magnetic torque. The 
eventual result is the enhanced transport of the outward vertical flux 
with the increase in the effective angular momentum transport in $z$ 
direction. This renders $v_{\varphi_2}$
to become negative at a lower $f$ and higher $\beta$ as compared to that for
$\alpha = 0.3$, as stated in last two lines of \S 4.3. 

In Fig. 7a we 
show the variation of the ratio of $\bar v_z$ and poloidal 
Alfven velocity $\bar v_{AP}$ [$= \bar B_P / \sqrt(4 \pi \bar \rho)$] 
in $z$ for different $\dot M$ and $\alpha$. With 
the increase in $\dot M$ as the system becomes less advective and 
more centrifugally/rotationally dominated, there is an increase in the ratio of $\bar v_z /{\bar v_{AP}}$. In 
contrast, with the increase in $\alpha$ for a specific $\dot M$ there 
is a sharp fall in the value of the above ratio. 
Figure 7b 
shows the variation of the ratio 
of $\bar v_z$ and net Alfven velocity $\bar v_{AT}$ 
[$= \bar B / \sqrt (4 \pi \bar \rho) $] in $z$ similar to that 
in Fig. 7a. Figure 7c depicts the profile of plasma $\beta_P$ 
[$= {\bar B}^2 / (8 \pi \bar \rho {\bar c}^2_s) $] in $z$ for 
different $\dot M$ and $\alpha$. 
We find that $\beta_P$ always increases steadily in the 
vertical direction, and its value gets strongly augmented with a small increase in 
$\alpha$, however, it always remains mostly below equipartition for 
all relevant $\dot M$ and $\alpha$. Moreover it is found that with the increase in $\dot M$, 
in general, $\beta_P$ decreases. 

In figures 8 and 9 we make a comparison of our magnetized accretion-outflow solutions with that of 
ADAF type solutions 
in [5]. In figures 8 and 9, we compare the radial profiles for radial 
velocity, orbital velocity, sound speed 
and density, respectively, obtained for our magnetized accretion-outflow solutions with that of 
ADAF type solutions, corresponding to different $\dot M$ and different $\alpha$. We found that the 
magnitude of radial velocity along 
$r$ that we obtained in our accretion-induced outflow is less as compared to that obtained in self-similar ADAF. On the other hand, 
the magnitude of orbital velocity along $r$ obtained in our case is higher as compared to that obtained in ADAF. Also there is a marginal 
decrease in the magnitude of sound speed or equivalently the temperature of the gas along $r$ obtained in our accretion-induced outflow 
as compared to that obtained in ADAF. However, the magnitude of density of the gas along $r$ in our accretion-induced outflow is 
found to be almost similar with that obtained in case of ADAF.  

\section{Discussion}

Observationally it is found that low/hard state of BHXRBs 
which are supposed to be powered by geometrically thick strongly advective 
sub-Eddington (presumably with $\dot M \, \lsim \, 10^{-3} \, \dot M_{\rm Edd}$) and consequently 
quasi-spherical and radiatively inefficient accretion flows (RIAFs), emanate strong 
outflows and relativistic jets. Outflows and jets are not 
observed in high/soft of BHXRBs which are powered by geometrically thin and optically 
thick standard Keplerian accretion disk. The physics of origin and launching of  outflows/jets 
in galactic BH systems (also called microquasars) is 
supposed to be similar with that corresponding to SMBHs in AGNs, as AGNs may be seen to be 
scaled up galactic BHs [37]. Geometrically 
thick advective accretion flow having substantial amount of gas 
pressure with strong advection, is more conducive to effuse and accelerate plasma 
in the vertical direction out of the inner accretion region. Although we do not aspire to explore the exact 
mechanism of launching and 
ejection of jets, however, it is generally conceived 
that the origin, launching and ejection of outflow and jet from 
the accretion flow is a magnetohydrodynamic process. In the present work we mainly focussed on accretion powered hydromagnetic 
outflows. 

Although the distinctive cause of the origin and launching of accretion powered hydromagnetic outflows 
is still inconclusive, however, it is certain that the dynamics of the outflowing 
matter should be intrinsically coupled to the accretion dynamics through the 
fundamental laws of conservation of (matter, momentum and energy) within 
the coupled accretion-induced outflow region, and should not 
be treated as dissimilar objects. 
Conservation laws are the most valuable foundation in physics, and
play a significant role in understanding astrophysical outflows and jets.
This is because the physical dynamics of the coupled inflow and outflow 
are essentially governed by the laws of conservation. The nature 
of the dynamical solutions in the accretion-outflow coupled region should then reflect upon the 
physical conditions/criteria  
to eject outflows. For the theoretical analysis of the 
accretion-outflow coupling one needs to be very thoughtful about the proper 
modelling of the system which essentially needs to solve a complete set of magnetohydrodynamic 
conservation equations in 2.5-dimension viscous, resistive, advective paradigm. In \S \RNum{2} we have endeavoured to 
describe a robust form of accretion-outflow coupled magnetohydrodynamic set of 
equations in viscous, resistive, advective paradigm, upholding all the conservation laws in 2.5-dimensional mean field 
MHD regime without any ad hoc proposition, where the dynamical 
flow variables vary in ($r,z$). The mean field approximation 
gives rise to the emergence of various turbulent correlation terms, where we restrict our study to first order turbulent correlation. 
Note that turbulent magnetic diffusivity and turbulent 
viscous term in induction and energy conservation equation arises only 
from the kinetic part of the turbulent stress tensor through the Reynolds stress.
In this work, we have assumed isotropic turbulence and also neglected 
the contribution of other turbulent stress tensors apart from $r\varphi$, which 
may be dynamically important. The contribution $r\varphi$ component 
would be dynamically more dominant as is responsible for the radial 
transport of angular momentum outwards. Vertical transport of angular momentum 
occurs mainly through large-scale magnetic stresses. Nonetheless, in future we 
would like to examine the possibility of their inclusion, as well as 
investigate the nature of the flow with anisotropic turbulence. 

The inflow and outflow are governed by eight coupled integro-partial 
differential MHD equations in the cylindrical geometry. Limited observational 
inputs put constraint on the boundary conditions as well 
as the scaling relation between accretion and the outflow. Owing to the fact 
that it is beyond the scope to have complete global numerical solutions of the said coupled partial differential MHD equations, motivated 
us to invoke necessary and proper quasi-analytical method to solve them. Ever since the work of [5], use 
of power law self-similarity in studying the accretion flow 
dynamics, especially the advection dominated accretion flows (ADAFs) to 
explain the nature of LLAGNs has become widely popular. Realistic strongly 
advective accretion flow preserves self-similarity reasonably well, within an appreciable region of 
the flow [5,38], and has been widely used to explain observational features in 
LLAGNs (see [39] for a review). Previously, self-similar methods had been indeed used to 
study outflow from the accretion disk on many 
occasions (see the references in \S \RNum{1} in introduction). It is being 
found from many studies that self-similarity holds approximately well in context to outflows 
[40,18,28,29,41].
Keeping the essence of power law 
self-similarity, we sought a generalized n$^{\textnormal{th}}$ degree 
polynomial expansion of all the dynamical variables in two dimensions, and 
solve the complete set of coupled integro-partial differential MHD conservation 
equations within the accretion-outflow coupled region self-consistently, in 2.5 dimensional 
viscous, resistive advective paradigm, where we have restricted up to 
order $[h(r)/r]^2$. 

In sections 4 and 5, we have analyzed the nature and the 
behaviour of our MHD solutions in advective accretion paradigm 
with $\dot M \, \lsim \, 10^{-2} \, \dot M_{\rm Edd}$. Although 
we have not intended to explore the physical 
mechanism of outflow/jet launching in the present study, the quasi-stationary dynamical solutions of the 
accretion-induced outflow carry the information about the physical conditions/criteria to propel matter vertically 
outwards out of the accretion-outflow region. We have mainly focused within the 
accretion-outflow coupled region where the flow is essentially 
bounded. We obtain solutions at a reduced vertical thickness irrespective of the nature of accretion 
paradigm which we have focused on. Magnetic 
field tends to compress or squeeze the accretion region by counterbalancing the 
thermal pressure gradient. We found that the large-scale poloidal component of the 
magnetic field is enhanced with the increase in the 
geometrical thickness of the accretion flow, consistently. With
the increase in $\dot M$ as $10^{-4} \dot M_{\rm Edd} << \dot M \, \lsim \, 10^{-2} \, \dot M_{\rm Edd}$, 
with the accretion flow becoming less advective 
and more centrifugally dominated with lesser geometrical thickness i.e., for the flow with moderate 
advection, there is a sharp fall in the value of poloidal component of the magnetic field, however with 
a strong enhancement in the
value of the toroidal component of the magnetic field and consequently the 
differential magnetic torque ($-r^2 \bar B_{\varphi} \bar B_z$). This term (differential magnetic torque) 
is responsible for the magnetic extraction of the 
angular momentum to magnetocentrifugally accelerate the outflowing plasma out of 
the radial accretion flow, and this predominantly determines the outward flow 
of the matter in a moderately advective accretion paradigm which is more 
centrifugally dominated. However, with the decrease in $\dot M$ as the flow becomes strongly advective 
($\dot M \, \lsim \, 10^{-3} \, \dot M_{\rm Edd}$) and geometrically more thick with strong gas pressure 
and inefficient cooling, despite in the 
decrease in the value of $-r^2 \bar B_{\varphi} \bar B_z$, a 
consistent increase in the $v_z$ occurs, indicating that the gas pressure gradient might play a more 
contributory role to lift the plasma vertically outwards with the 
help of magnetic forces. The plasma in the accretion flow can be lifted outwards and ejected, only if 
some physical process can overcome the effect of the inward vertical force due 
to the central gravity. The dynamical behaviour of the solutions 
indicate that in the advective paradigm both magnetocentrifugal acceleration and thermal pressure gradient  
along with the magnetic forces, will help in lifting and accelerating the plasma vertically outwards, and 
the gas material will diffuse outwards across magnetic field lines. However, the 
effective contribution of either magnetocentrifugal acceleration 
or thermal pressure gradient to lift the plasma vertically outwards depends 
on the degree to which the flow is advective. In fact, with the increase in mass accretion rate as the flow tends to become less 
advective and more centrifugally dominated with lesser 
geometrical thickness, in general, the efficacy of the disk to eject outflows diminishes. In paper \RNum{2} (in preparation), we 
have quantitatively demonstrated this aspect with the increase in $\dot M$ from $10^{-4} \dot M_{\rm Edd}$ to $10^{-2} \dot M_{\rm Edd}$; the 
accretion flow with $\dot M \sim 10^{-2} \dot M_{\rm Edd}$, least conducive to eject outflows.

We obtain dynamical solutions in accretion-outflow coupled region only 
at high turbulent diffusive parameter $\alpha \, (\gsim \, 0.3)$. The accretion-induced 
outflow solutions have a profound dependence on turbulent diffusive parameter $\alpha$. It is 
being interestingly found from our solutions that the enhancement in $\alpha$ renders the accretion-induced outflow 
region to attain a greater geometrical thickness. Consequently, the poloidal component of magnetic field   
$\bar B_P$, as well as the differential magnetic torque ($-r^2 \bar B_{\varphi} \bar B_z$) get strongly 
augmented, enhancing the transport of vertical flux outwards. Also the plasma beta 
$\beta_P$ increases steadily in the vertical direction, and 
its value gets strongly augmented with a small increase in $\alpha$, 
however, it always remains mostly below equipartition within the 
accretion-outflow coupled region. Although we expect the accretion flow to have 
a large $\alpha$ owing to advective nature of the flow, however, the values of $\alpha$ 
that we have obtained in our solution may have been slightly overestimated. 
Nonetheless, it is indeed being found from previous works [5,6,38,42]
that strongly advective accretion flows, in general, are mostly plausible for large values of $\alpha$ 
($\alpha \, \gsim \, 0.1$). Gu and Lu [43] also shown that the transition from an outer 
geometrically thin Keplerian disk to an advection-dominated accretion flow 
is possible for $\alpha > 0.5$. McKinney and Narayan [29], in their GRMHD simulation of disk-outflow model, 
found to have large turbulent viscosity parameter in the accretion disk in the vicinity of the BH. Further,  
King, Pringle and Livio [44] suggested a typical range of $\alpha \sim (0.1 - 0.4)$ from observational evidence. 

One of the important approximation we have used in our study is to treat 
the scale-height of accretion-outflow coupled region as a parameter. In 
reality the physical conditions to launch outflow would consistently determine 
the vertical height of the inflow-outflow surface, from where the outflow 
decouples from the accretion region. Moreover, in our 
study we have neglected the effect of spin of the BH in 
the accretion dynamics and its subsequent impact on the outflow, as 
the self-similar technique can only be used in the Newtonian approximation. 
This restricts us of using this method to investigate the physical behaviour 
of the system in the extreme vicinity of the BH, where 
general relativistic effects are indispensable. Although power law 
self-similarity is an analytical approximation, and the quantitative 
feature of the solutions may have been either overestimated or undervalued, 
the dynamical solutions 
show consistent and predictable behaviour, and 
do exhibit many physical insights on the nature of the accretion-induced 
outflow, as well reflect upon the relevant physical conditions to propel 
and eject plasma out of the accretion flow. Power law self-similarity, thus, seems to be 
reasonably good approximation, within the 
accretion-outflow coupled region. 

Nonetheless, more extensive study is required to understand the definitive 
criteria or condition in launching accretion powered outflows and jets. 
BH spin is a very important aspect that needs to be 
incorporated in the conservation equations atleast through the use of pseudo-Newtonian potentials (e.g., [45]), 
while understanding accretion powered outflow dynamics or the correlated dynamics of accretion 
and outflow. It is found from the work of [10] that the spin of the BH significantly influences 
accretion powered outflows/jets; for a rapidly rotating BH, the outflow power increases by $\sim$ two orders in magnitude. 
Moreover, it is also 
essential that to have complete and a more realistic understanding of the dynamics of the accretion-outflow coupled region, a 
global numerical solution of such a system in advective paradigm should be performed. 
Also, explicit 
inclusion of cooling/radiative processes is required for the completeness of energy conservation. The relevant dynamical 
solutions at the accretion-outflow coupled surface would then necessarily act as boundary conditions at the base of the jet.
A more definitive understanding of the criteria to launch accretion powered outflows and jets, thus, requires a complete 
2.5-dimensional viscous, resistive, advective global MHD numerical solution with the inclusion of BH spin, which 
is left for future work. In a subsequent work [paper \RNum{2} (in preparation)] we will investigate in detail, the energetics of the magnetized 
accretion-induced outflows and study the spectral behaviour of accretion powered sources.


\appendix
\section{}


The integro-differential continuity equation (\ref{7}), after substitution 
of the flow variables in the n$^{\textnormal{th}}$ polynomial order is given by  
$$
\sum_{n=0}^{\infty} \Bigl[\frac{1}{2n+1} \sum_{m=0}^{n} \rho_{2 (n-m)} v_{r2m} + \frac{1}{e+c-2n+2} \sum_{m=0}^{n} \rho_{2 (n-m)} v_{z (2m+1)} \Bigr] \,  \Big(\frac{h}{r}\Big)^{2n+1}= - \frac{\dot M}{4 \pi} 
\eqno(A1)
$$

Similarly the polynomial expansion of all the other height-integrated 
MHD equations are done, however we do not furnish all of them here 
as the structure 
of the equations are huge. As an example we show it for the radial momentum 
balance equation (\ref{9}). \\
$$
\Biggl[
\sum_{n=0}^{\infty} \sum_{m=0}^{n} \sum_{l=0}^{m}\, (a-2l) \,\rho_{2 (n-m)} 
\, v_{r2 (m-l)} \, v_{r2l} \, 
- \, \sum_{n=0}^{\infty} \sum_{m=0}^{n} \sum_{l=0}^{m} \, \rho_{2 (n-m)}\, v_{\varphi 2 (m-l)} \, v_{\varphi 2l} \, 
$$
$$
\, + \, \sum_{n=1}^{\infty} \sum_{m=1}^{n} \sum_{l=1}^{m} \, 2l \, \rho_{2 (n-m)} \, v_{z [2 (m-l)+1 ]} \, v_{r 2l} \, + \, 
G M \sum_{n=0}^{\infty} \sum_{m=0}^{n} {-3/2 \choose m} \, \rho_{2 (n-m)} \,
$$
$$
\, + \, \sum_{n=0}^{\infty} \sum_{m=0}^{n} \sum_{l=0}^{m} 
\, (e-2n +2d) \, \rho_{2 (n-m)} \, c_{s2 (m-l)} \, c_{s2l} \, 
+ \, \frac{1}{4 \pi} \biggl[
\sum_{n=0}^{\infty} \sum_{m=0}^{n} \, (j-2m+1) \, B_{\varphi 2(n-m)} \, B_{\varphi 2m} 
\, 
$$
$$ 
\, + \,  \sum_{n=1}^{\infty} \sum_{m=1}^{n} \, [k-2(m-1)] \, B_{z [2(n-m)+1]} \, 
B_{z [2(m-1)+1]} \, - \, \sum_{n=1}^{\infty} \sum_{m=1}^{n} \, 2m \, 
B_{z [2(n-m)+1]} \, B_{r2m} \biggr] \, \Biggr] \, 
\frac{1}{2n+1} \, \Bigl(\frac{h}{r}\Bigr)^{2n} = 0.
\eqno(A2)
$$

\end{document}